\documentclass{elsart}

\usepackage{epsfig}

\newcommand{\be}{\begin{equation}}
\newcommand{\ee}{\end{equation}}
\newcommand{\bea}{\begin{eqnarray}}
\newcommand{\eea}{\end{eqnarray}}

 \def\bean{\begin{eqnarray*}}
 \def\eean{\end{eqnarray*}}

 \def\l{\left}
 \def\r{\right}
 \def\Im{{\rm Im}}
 \def\Re{{\rm Re}}
 \def\bm#1{\mbox{\boldmath$#1$}}
 \def\gsim{\mathrel{\rlap{\lower0.2em\hbox{$\sim$}}\raise0.2em\hbox{$>$}}}
 \def\ksim{\mathrel{\rlap{\lower0.2em\hbox{$\sim$}}\raise0.2em\hbox{$<$}}}

\begin{document}

\begin{frontmatter}
\title{Dynamical quasiparticles properties and effective interactions in the sQGP}
\author[unig]{W. Cassing\corauthref{cor1}}
\ead{Wolfgang.Cassing@theo.physik.uni-giessen.de}
\corauth[cor1]{corresponding author}
\address[unig]{Institut f\"ur Theoretische Physik, %
  Universit\"at Giessen, %
  Heinrich--Buff--Ring 16, %
  D--35392 Giessen, %
  Germany}

\begin{abstract}
  Dynamical quasiparticle properties are determined from lattice
  QCD along the line of the Peshier model for the running strong
  coupling constant in case of three light flavors. By separating
  time-like and space-like quantities in the  number
  density and energy density the effective degrees of freedom in
  the gluon and quark sector may be specified from the time-like densities.
  The space-like parts of the energy densities are identified with
  interaction energy (or potential energy) densities.
  By using the time-like parton densities (or scalar densities)
  as independent degrees of freedom -
  instead of the temperature $T$ and chemical potential $\mu_q$
  as Lagrange parameters -
  variations of the potential energy densities with respect to the time-like
  gluon and/or fermion densities lead to effective
  mean-fields for time-like gluons and quarks as well as to effective
  gluon-gluon, quark-gluon and quark-quark (quark-antiquark) interactions.
  The latter dynamical quantities are found to be approximately
  independent on the quark chemical potential $\mu_q$ and thus well
  suited for an inplementation in off-shell parton transport
  approaches. Results from the dynamical quasiparticle model (DQPM)
   in case of two dynamical light
  quark   flavors are compared to lattice QCD calculations for the net
  quark density $\rho_q(T,\mu_q)$ as well as for the
  'back-to-back' differential dilepton production rate by $q-{\bar
  q}$ annihilation. The DQPM is found
  to pass the independent tests.
\end{abstract}


\end{frontmatter}


\section{Introduction}
The 'Big Bang' scenario implies that in the first micro-seconds of
the universe the entire system has emerged from a partonic system
of quarks, antiquarks and gluons -- a quark-gluon plasma (QGP) --
to color neutral hadronic matter consisting of interacting
hadronic states (and resonances) in which the partonic degrees of
freedom are confined. The nature of confinement and the dynamics
of this phase transition has motivated a large community for
several decades (cf.\ \cite{Tannenbaum,Jacobs,QM01} and Refs.\
therein). Early concepts of the QGP were guided by the idea of a
weakly interacting system of partons since the entropy density $s$
and energy density $\epsilon$ were found in lattice QCD to be
close to the Stefan Boltzmann (SB) limit for a relativistic
noninteracting system \cite{Karsch}. However, experimental
observations at the Relativistic Heavy Ion Collider (RHIC)
indicated that the new medium created in ultrarelativistic Au+Au
collisions was interacting more strongly than hadronic matter and
consequently this notion had to be given up. Moreover, in line
with earlier theoretical studies in Refs.
\cite{Thoma,Andre,Shuryak} the medium showed phenomena of an
almost perfect liquid of partons \cite{STARS,Miklos3} as extracted
from the strong radial expansion and elliptic flow of hadrons as
well the scaling of the elliptic flow with parton number {\it
etc}. All the latter collective observables have been severely
underestimated in conventional string/hadron transport models
\cite{Cassing03,Brat04,Cassing04} whereas hydrodynamical
approaches did quite well in describing (at midrapidity) the
collective properties of the medium generated during the early
times for low and moderate transverse momenta \cite{Heinz,Bass2}.
The question about the constituents and properties of this QGP
liquid is discussed controversely in the literature (cf. Refs.
\cite{GerryEd,GerryRho,Eddi}) and practically no dynamical
concepts are available to describe the dynamical freezeout of
partons to color neutral hadrons that are finally observed
experimentally. Since the partonic system appears to interact more
strongly than even hadronic systems the notation strong QGP (sQGP)
has been introduced in order to distinguish from the dynamics
known from perturbative QCD (pQCD).

Lattice QCD (lQCD) calculations provide some guidance to the
thermodynamic properties of the partonic medium close to the
transition at a critical temperature $T_c$ up to a few times
$T_c$, but lQCD calculations for transport coefficients presently
are not accurate enough \cite{lattice2} to allow for firm
conclusions. Furthermore, it is not clear whether the partonic
system really reaches thermal and chemical equilibrium in
ultrarelativistic nucleus-nucleus collisions \cite{ZhangKo} such that
nonequilibrium models are needed to trace the entire collision
history.  The available string/hadron transport models
\cite{Cass99,URQMD1,URQMD2} partly fail - as pointed out above -
nor do partonic cascade simulations
\cite{Geiger,Zhang,Molnar,Bass} (propagating massless partons)
sufficiently describe the reaction dynamics when employing cross
sections from perturbative QCD. Some models, e.g. the
Multiphase Transport Model AMPT \cite{AMPT}, employ strong
enhancement factors for the cross sections, however, use only
on-shell massless partons in the partonic phase as in Ref.
\cite{Zhang}. The same problem comes about in the parton cascade
model of Ref. \cite{Carsten} where additional 2$ \leftrightarrow$
3 processes like $gg \leftrightarrow ggg$ are incorporated but
massless partons are considered.

On the other hand it is well known that strongly interacting
quantum systems require descriptions in terms of propagators $D$
with sizeable selfenergies $\Pi$ for the relevant degrees of
freedom. Whereas the real part of the selfenergies can be related
to  mean-field potentials, the imaginary parts of $\Pi$ provide
information about the lifetime and/or reaction rate of time-like
'particles' \cite{Andre}. In principle, off-shell transport
equations are available in the literature
\cite{Juchem,Sascha1,Leo}, but have been applied only to dynamical
problems where the width of the quasiparticles stays moderate with
respect to the pole mass \cite{Laura}. On the other hand, the
studies of Peshier \cite{Andre04,Andre05} indicate that the
effective degrees of freedom in a partonic phase should have a
width $\gamma$  in the order of the pole mass $M$ already slightly
above $T_c$. This opens up the problem how to interpret/deal with
the space-like part of the distribution functions and how to
'pro\-pagate' effective degrees in space-time in equilibrium as
well as out of equilibrium.

 Some elementary steps in answering these
questions have been presented in a preceeding work
\cite{Cassing06} where the pure Yang-Mills sector of QCD has been
addressed in the Dynamical QuasiParticle model (DQPM) of Peshier
\cite{Andre04,Andre05}. In the latter work it could be
demonstrated that the DQPM allows for a rather simple and
transparent interpretation of QCD thermodynamics and leads to
effective strongly interacting gluonic degrees of freedom that may
be propagated in off-shell transport approaches as e.g. in Ref.
\cite{PHSD}. The present study is an extension of the work in Ref.
\cite{Cassing06} to three light quark flavors ($q= u,d,s$) and to
finite quark chemical potential $\mu_q$ in order to determine the
effective partonic degrees of freedom, their mean-fields and
interactions for a practical implementation in the off-shell
parton transport model PHSD\footnote{Parton-Hadron-String
Dynamics} \cite{PHSD}. First results of the PHSD approach for
dynamical phase trajectories in central Au+Au collisions have been
presented in Ref. \cite{Arsene}.

The outline of the paper is as follows: After a short
recapitulation of the dynamical quasiparticle model with dynamical
quarks in Section 2 novel results on the space-like and time-like
parts of observables are presented that allow for a transparent
physical interpretation. In Section 3 we will examine derivatives
of the space-like part of the quasiparticle energy density with
respect to the time-like  densities which provides
information on partonic mean-fields and their effective
interaction strength. An extension of the model to finite quark
chemical potentials is given in Section 4 including a comparison
of the net quark density $\rho_q(T, \mu_q)$ with lQCD calculations
in the two flavor case. In Section 5 the 'back-to-back' dilepton
production rate is calculated in the DQPM and compared to
respective lQCD calculations again for two dynamical quark
flavors. A summary and short discussion closes this work in
Section 6.

\section{Off-shell elements in the DQPM}
\subsection{Reminder of the DQPM}
The dynamical quasiparticle model  adopted here goes back to
Peshier \cite{Andre04,Andre05} and starts with the entropy density
 in the quasiparticle limit ~\cite{Andre05,R38,R39,R40},
\be   \label{sdqp} \hspace{0.5cm}
  s^{dqp}
  =
  - d_g \!\int\!\!\frac{d \omega}{2 \pi} \frac{d^3p}{(2 \pi)^3}
  \frac{\partial n_B}{\partial T}
   \l( \Im\ln(-\Delta^{-1}) + \Im\Pi\,\Re\Delta \r) \ee $$
   - d_q \!\int\!\!\frac{d \omega}{2 \pi} \frac{d^3p}{(2 \pi)^3}
  \frac{\partial n_F((\omega-\mu_q)/T)}{\partial T}
   \l( \Im\ln(-S_q^{-1}) + \Im\Sigma_q\,\Re S_q \r)
   \!, $$
$$
   - d_{\bar q} \!\int\!\!\frac{d \omega}{2 \pi} \frac{d^3p}{(2 \pi)^3}
  \frac{\partial n_F((\omega+\mu_q)/T)}{\partial T}
   \l( \Im\ln(-S_{\bar q}^{-1}) + \Im\Sigma_{\bar q}\,\Re S_{\bar q} \r)
   \!, $$

\noindent where $n_B(\omega/T) = (\exp(\omega/T)-1)^{-1}$ and
$n_F((\omega-\mu_q)/T) = (\exp((\omega-\mu_q)/T)+1)^{-1}$ denote the
Bose and Fermi distribution functions, respectively, while $\Delta
=(P^2-\Pi)^{-1}$, $S_q = (P^2-\Sigma_q)^{-1}$ and $S_{\bar q} =
(P^2-\Sigma_{\bar q})^{-1}$ stand for scalar quasiparticle
propagators of gluons $g$, quarks $q$ and antiquarks ${\bar q}$. The degeneracy
factor for gluons is $d_g= 2 (N_c^2-1)$ = 16 while for quarks $q$
and antiquarks ${\bar q}$ we get $d_q = d_{\bar q} = 2 N_c N_f
7/8$ = 15.25 for three flavors $N_f$. In (\ref{sdqp}) $\Pi$ and
$\Sigma = \Sigma_q \approx \Sigma_{\bar q}$ denote the
quasiparticle selfenergies. In principle, $\Pi$ as well as
$\Delta$ are Lorentz tensors and should be evaluated in a
nonperturbative framework. The DQPM treats these degrees of
freedom as independent scalar fields with a scalar selfenergy
$\Pi$. In case of the fermions $S_q, S_{\bar q}$ and $\Sigma_q,
\Sigma_{\bar q}$ (for $q$ and ${\bar q}$) have Lorentz scalar and
vector contributions but only scalar terms are kept in
(\ref{sdqp}) for simplicity which are assumed to be identical for
quarks and antiquarks. Note that one has to treat quarks and
antiquarks separately in (\ref{sdqp}) as their abundance differs
at finite quark chemical potential $\mu_q$.

Since the nonperturbative evaluation of the propagators and
selfenergies is a formidable task (and presently not solved) a
more practical procedure is to use physically motivated {\em
Ans\"atze} with Lorentzian spectral functions for quarks\footnote{In the following
the abbreviation is used that 'quarks' denote quarks and antiquarks
if not specified explicitly.}
and gluons,
\be
 \rho(\omega)
 =
 \frac\gamma{ E} \l(
   \frac1{(\omega-E)^2+\gamma^2} - \frac1{(\omega+E)^2+\gamma^2}
 \r) ,
 \label{eq:rho}
\ee and to fit the few parameters to results from lQCD. With the
convention $E^2(\bm p) = \bm p^2+M^2-\gamma^2$, the parameters
$M^2$ and $\gamma$ are directly related to the real and imaginary
parts of the corresponding (retarded) self-energy, e.g. $\Pi =
M^2-2i\gamma\omega$ in case of the 'scalar' gluons.

Following  \cite{Andre05} the quasiparticle mass (squared) for
gluons is assumed to be given by the thermal mass in the
asymptotic high-momentum regime, i.e.
\be
 M^2(T) = \frac{g^2}{6} \left( (N_c + \frac{1}{2}N_f)\, T^2
 + \frac{N_c}{2} \sum_q \frac{\mu_q^2}{\pi^2}
 \right) \, ,
 \label{eq:M2} \ee
and for quarks (assuming vanishing constituent masses) as,
\be
m^2(T) = \frac{N_c^2-1}{8 N_c}\, g^2 \left( T^2 +
\frac{\mu_q^2}{\pi^2} \right) \, ,\label{eq:M2b} \ee with a
running coupling (squared),
\be
 g^2(T/T_c) = \frac{48\pi^2}{(11N_c - 2 N_f)  \ln(\lambda^2(T/T_c-T_s/T_c)^2}\ ,
 \label{eq:g2}
\ee which permits for an enhancement near $T_c$
\cite{pQP,Rossend,Rafelski}. Here $N_c = 3$ stands for the number
of colors while $N_f$ denotes the number of flavors and $\mu_q$
the quark chemical potentials. The parameters $\lambda = 2.42$ and
$T_s/T_c = 0.46$ are adopted from \cite{Andre05}. As demonstrated
in Fig. 1 of Ref. \cite{Cassing06} this functional form for the
strong coupling $\alpha_s = g^2/(4\pi)$ is in accordance with the
lQCD calculations of Ref. \cite{Bielefeld} for the long range part
of the $q - \bar{q}$ potential.

The width  for gluons and quarks (for $\mu_q = 0$) is adopted in
the form \cite{Pisar89LebedS}
\be
  \gamma_g(T)
  =
  N_c \frac{g^2 T}{8 \pi} \,  \ln\frac{2c}{g^2} \, , \hspace{2cm}
    \gamma_q(T)
  =
  \frac{N_c^2-1}{2 N_c} \frac{g^2 T}{8 \pi} \,  \ln\frac{2c}{g^2}
  \,.
 \label{eq:gamma}
\ee where $c=14.4$ (from Ref. \cite{Andre}) is related to a
magnetic cut-off. Note that in case of vanishing number of flavors
$N_f=0$ the expressions for the masses and the coupling reduce to
those employed for the pure Yang-Mills sector
\cite{Andre05,Cassing06}.

The physical processes contributing to the width $\gamma_g$ are
both $gg \leftrightarrow gg$, $gq \leftrightarrow gq$ scattering
as well as splitting and fusion reactions $gg \leftrightarrow g$,
$gg \leftrightarrow ggg$, $ggg \leftrightarrow gggg$ or $g
\leftrightarrow q \bar{q}$ etc. On the fermion side elastic
fermion-fermion scattering $pp \leftrightarrow pp$, where $p$
stands for a quark $q$ or antiquark $\bar{q}$, fermion-gluon
scattering $pg \leftrightarrow pg$, gluon bremsstrahlung $pp
\leftrightarrow pp+g$ or quark-antiquark fusion $q \bar{q}
\leftrightarrow g$ etc. emerge. Note, however, that the explicit
form of (\ref{eq:gamma}) is derived for hard two-body scatterings
only. It is worth to point out that the ratio of the masses to
their widths $ \sim g \ln(2c/g^2)$ approaches zero only
asymptotically for $T \rightarrow \infty$ such that the width of
the quasiparticles is comparable to the pole mass slightly above
$T_c$ up to all terrestrial energy scales.

Within the DQPM the real and imaginary parts of the propagators
$\Delta$ and $S$ now are fixed and the entropy density
(\ref{sdqp}) can be evaluated numerically once the free parameters
in (\ref{eq:g2}) are determined. In the following we will assume 3
light quark flavors $N_f = 3$. Since the presently available
unquenched lQCD calculations (for three flavors) for the entropy
density are still accompanied with rather large error bars the
parameters of the DQPM are taken the same as in the pure
Yang-Mills sector:  $\lambda = 2.42$, $T_s/T_c= 0.46$ as
determined in Ref. \cite{Andre}. This is legitimate since an
approximate scaling of thermodynamic quantities from lQCD is
observed when dividing by the number of degrees of freedom and
scaling by the individual critical temperature $T_c$ which is a
function of the different number of parton species \cite{Karsch5}.
However, these parameters will have to be refitted once more
accurate 'lattice data' become available.

\begin{figure}[htb!]
  \vspace{0.5cm}
    \includegraphics[width=11.5cm]{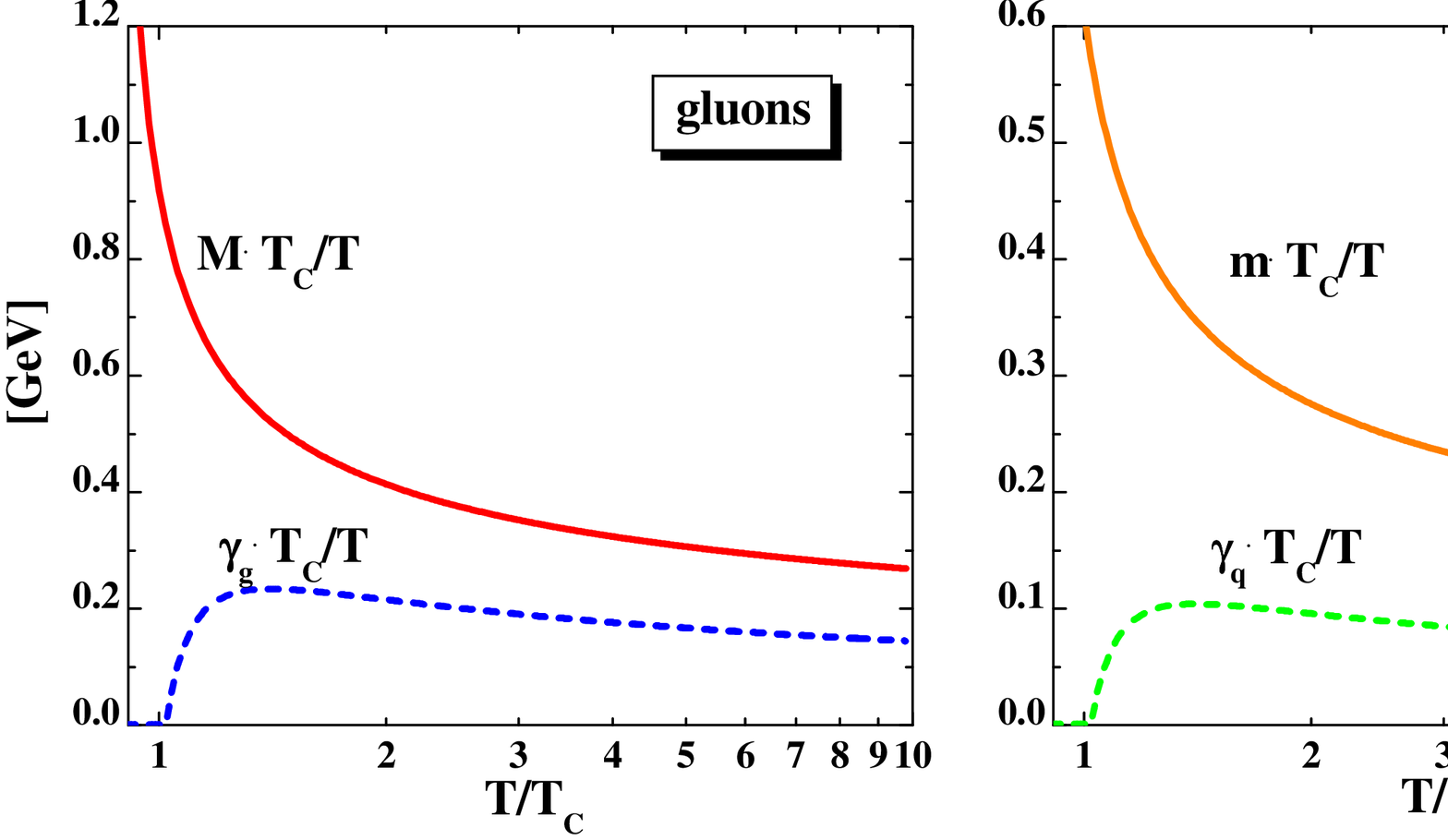}
    \caption{ The mass $M$ (solid red line) and width $\gamma_g$ (dashed blue line)
    for gluons (l.h.s.) and the mass $m$ (solid orange line) and width
    $\gamma_q$ (dashed green line) for
     quarks (r.h.s.) as a function of $T/T_c$ in the DQPM
     for $\lambda = 2.42$, $T_s/T_c= 0.46$, and $c=$ 14.4 (for $\mu_q = 0$).
     All quantities have been multiplied
     by the dimensionless factor $T_c/T$.}
    \label{fig1}
\end{figure}

The resulting values for the gluon and quark masses - multiplied
by $T_c/T$ - are displayed in Fig. \ref{fig1} (for $\mu_q=0$) by
the solid lines while the gluon and quark width ($\gamma_g,
\gamma_q)$ - multiplied by $T_c/T$ - are displayed in terms of the
dashed lines as a function of $T/T_c$.  The actual numbers are
found to be quite similar as in the pure Yang Mills case
\cite{Andre,Andre04,Andre05} and imply very 'broad' quasiparticles
already slightly above $T_c$. For $\mu_q = 0$ the ratio
$\gamma_q/\gamma_g$ =4/9 is the same as for the ratio of the
squared masses $m^2/M^2 = 4/9$ and reflects the ratio of the
Casimir eigenvalues in color space. Consequently the ratio of the
width to the pole mass is smaller for quarks (antiquarks) than for
gluons in the whole temperature range.

In order to fix the scale $T_c$, which is not specified so far,
one may directly address unquenched lQCD calculations (for 3 light
flavors). However, here the situation is presently controversal
between different groups (cf. Refs. \cite{xx1,xx2} and the
discussion therein). An alternative way is to calculate the
pressure $P$ from the thermodynamical relation,
\be
\label{pressure} s =\frac{\partial P}{dT} \ , \ee by integration
of the entropy density $s$ over $T$, where one may tacitly
identify the 'full' entropy density $s$ with the quasiparticle
entropy density $s^{dqp}$ (\ref{sdqp}). Since for $T < T_c$ the
DQPM entropy density drops to zero (with decreasing $T$) due to
the high quasiparticle masses and the width $\gamma$ vanishes as
well (cf. Fig. \ref{fig1}) the integration constant may be assumed
to be zero in the DQPM which focusses on the quasiparticle
properties above $T_c$.

\begin{figure}[htb!]
  \vspace{0.9cm}
    \includegraphics[width=10.0cm]{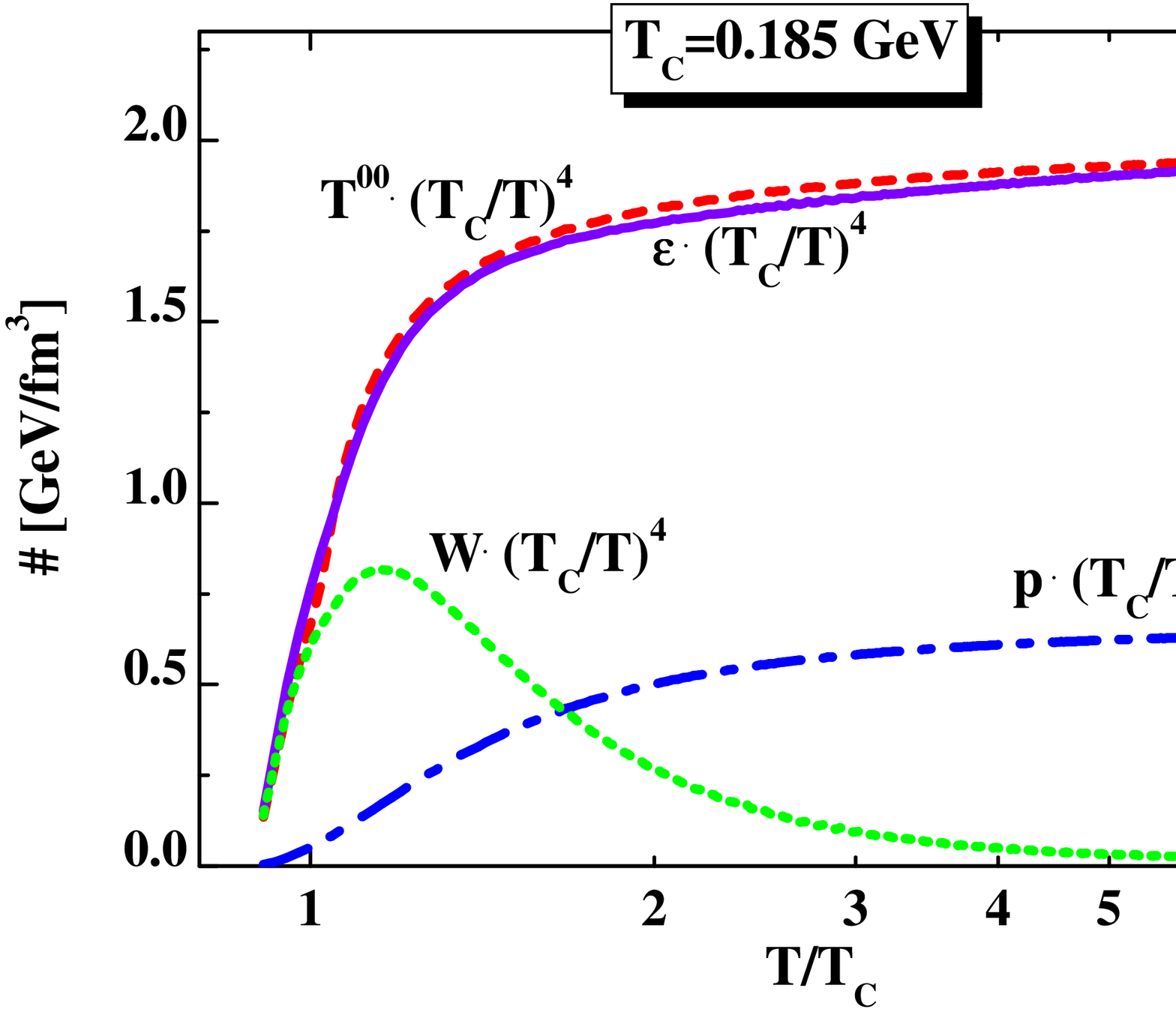}
    \caption{The DQPM results for $\epsilon (T/T_c) (T_c/T)^4$  (solid violet line),
    $P(T/T_c) (T_c/T)^4$ (blue
dash-dotted line) and the interaction measure $W(T/T_c) (T_c/T)^4$
(\ref{wint}) (green dotted line). Note the logarithmic scale in
$T/T_c$. The energy density $\epsilon$ (\ref{eps}) practically
coincides with the quasiparticle energy density $T^{00}$ from
(\ref{ent}) (dashed red line). }
    \label{fig2}
\end{figure}

\vspace{1.0cm}

The energy density $\epsilon$ then follows from the
thermodynamical relation \cite{pQP,Peshi} \be \label{eps} \epsilon
= T s -P \ee and thus is also fixed by the entropy $s(T)$ as well
as the interaction measure \be \label{wint} W(T): = \epsilon(T) -
3P(T) = Ts - 4 P \ee that vanishes for massless and noninteracting
degrees of freedom.

The actual results for  $\epsilon \cdot (T_c/T)^4$ are displayed
in Fig. \ref{fig2} (solid violet line), $P \cdot (T_c/T)^4$ (blue
dash-dotted line) as well as the interaction measure $W \cdot
(T_c/T)^4$ (\ref{wint}) (green dotted line) and show the typical
pattern from lQCD calculations \cite{Karsch5}. The scale $T_c$ may
now be fixed (estimated) by requiring that the critical energy
density $\epsilon(T_c)$ is roughly the same for the pure
Yang-Mills case as for the full theory with dynamical quarks in
line with the approximate scaling of lQCD \cite{Karsch5}. Since
$\epsilon (T_c)$ in the Yang Mills sector is about 1 GeV/fm$^3$
(cf. \cite{Cassing06}) the scaled energy density $\epsilon(T/T_c)$
from Fig. \ref{fig2} can be employed to fix the critical
temperature $T_c \approx$ 0.185 GeV. This leads to the 'thumb
rule' $\epsilon \approx 2\  (T/T_c)^4$ [GeV/fm$^3$] for $T > 1.2
T_c$ which is roughly fulfilled according to Fig. \ref{fig2}.

\begin{figure}[htb!]
  \vspace{0.9cm}
    \includegraphics[width=10.0cm]{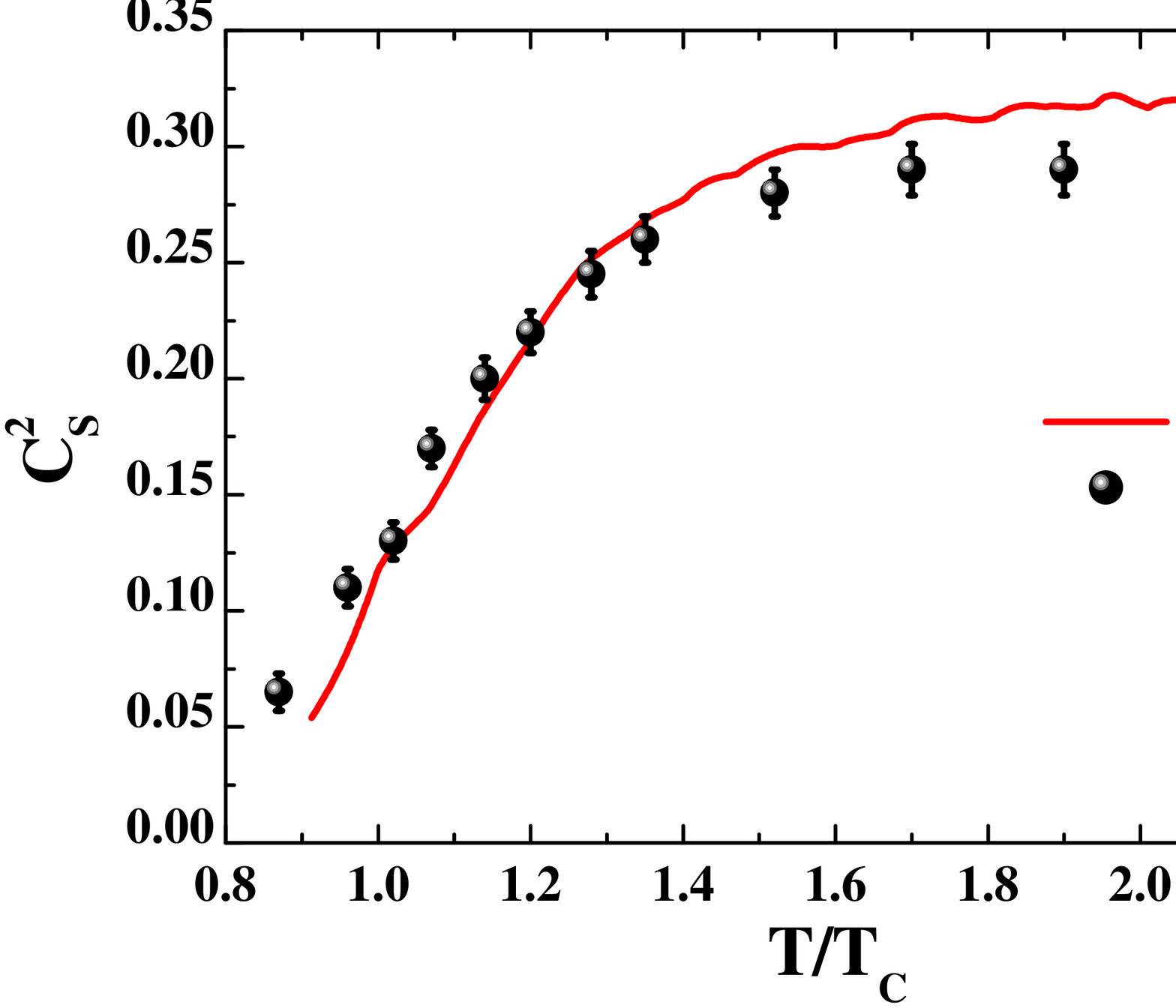}
    \caption{The DQPM results for the sound velocity (squared) (\ref{sound})
    as a function of $T/T_c$ in comparison to the lQCD results from Ref.
    \cite{Fodor5}.}
    \label{fig2b}
\end{figure}

\vspace{1.0cm}

A first test of the DQPM in comparison to lQCD calculations is
given for the sound velocity (squared), \be \label{sound} c_s^2 =
\frac{d P}{d \epsilon} \ , \ee which does not depend on the
absolute (uncertain) scale of $T_c$. This comparison is shown in
Fig. \ref{fig2b} by the solid line where the lQCD results have
been taken from Ref. \cite{Fodor5} and correspond to $N_f = 2+1$
with $N_t = 6$. The DQPM is seen to reproduce the drop in $c_s^2$
close to $T_c$ within errorbars and to reach the asymptotic value
$c_s^2$ = 1/3 approximately for $T > 2 \ T_c$. This comparison,
however, has to be taken with some care since the DQPM assumes
massless current quarks whereas the lQCD calculations employ
finite quark masses.

\subsection{Time-like and space-like quantities}
For the further analysis of the DQPM  it is useful to introduce
the shorthand notations (extending \cite{Andre,Cassing06}) \be
\label{conv} \hspace{1.5cm}
 {\rm \tilde Tr}^{\pm}_g \cdots
 =
 d_g\!\int\!\!\frac{d \omega}{2 \pi} \frac{d^3p}{(2 \pi)^3}\,
 2\omega\, \rho_g(\omega)\, \Theta(\omega) \, n_B(\omega/T) \ \Theta(\pm P^2) \, \cdots
 \,\ee
$$   {\rm \tilde Tr}^{\pm}_q \cdots
 =
 d_q\!\int\!\!\frac{d \omega}{2 \pi} \frac{d^3p}{(2 \pi)^3}\,
 2\omega\, \rho_q(\omega)\, \Theta(\omega) \, n_F((\omega-\mu_q)/T) \ \Theta(\pm P^2) \, \cdots
 \,$$ $$  {\rm \tilde Tr}^{\pm}_{\bar q} \cdots =
 d_{\bar q}\!\int\!\!\frac{d \omega}{2 \pi} \frac{d^3p}{(2 \pi)^3}\,
 2\omega\, \rho_{\bar q}(\omega)\, \Theta(\omega) \, n_F((\omega+\mu_q)/T) \ \Theta(\pm P^2) \, \cdots
$$

\noindent
 with $P^2= \omega^2-{\bf p}^2$ denoting the invariant mass
squared.  The quark and
antiquark degrees of freedom, i.e. 2$d_q$ = 31.5, are consequently
by roughly a factor of two more abundant than the gluonic degrees
of freedom. The $\Theta(\pm P^2)$ function in (\ref{conv})
separates time-like quantities from space-like quantities and can
be inserted for any observable of interest. Note, however, that
not all space-like quantities have a direct physical
interpretation.

We note in passing that the entropy density (\ref{sdqp}) is
dominated by the time-like contributions for quarks and gluons and
shows only minor space-like parts (cf. Refs.
\cite{Andre05,Cassing06}). Furthermore, the entropy density from
the DQPM is only 10 - 15 $\%$ smaller than the Stefan Boltzmann
entropy density $s_{SB}$ for $T > 2\  T_c$ as in case of lQCD
\cite{Karsch}. Since these results provide no novel information an
explicit representation is discarded.

Further quantities of interest are the 'quasiparticle densities'
\be
   N^{\pm}_g (T) = {\rm {\tilde Tr}^{\pm}_g }\ 1, \hspace{1cm}
   N^{\pm}_q (T) = {\rm {\tilde Tr^{\pm}_q }}\ 1, \hspace{1cm}
   N^{\pm}_{\bar q} (T) = {\rm {\tilde Tr^{\pm}_{\bar q} }}\ 1,
   \label{eq: N+}
\ee that correspond to the time-like (+) and space-like (-) parts
of the integrated distribution functions. Note that only the
time-like integrals over space have a particle number
interpretation. In QED this corresponds e.g. to time-like photons
($\gamma^*$) which are virtual in intermediate processes but may
also be seen asymptotically by dileptons (e.g. $e^+ e^-$ pairs)
due to the decay $\gamma^* \rightarrow e^+e^- (\mu^+ + \mu^-)$
\cite{Cass99}.

Scalar densities for quarks and gluons - only defined in the
time-like sector - are given by \be \label{scalar} N^s_g(T) = {\rm
{\tilde Tr}^+_g }\ \left( \frac{\sqrt{P^2}}{\omega} \right), \,
\hspace{0.3cm}  N^s_q(T) = {\rm{\tilde Tr^+_q }}\ \left(
\frac{\sqrt{P^2}}{\omega} \right), \, \hspace{0.3cm}  N^s_{\bar
q}(T) = {\rm{\tilde Tr^+_{\bar q} }}\ \left(
\frac{\sqrt{P^2}}{\omega} \right) \, \ee and have the virtue of
being Lorentz invariant.

\begin{figure}[htb!]
  \vspace{0.9cm}
   \includegraphics[width=11.5cm]{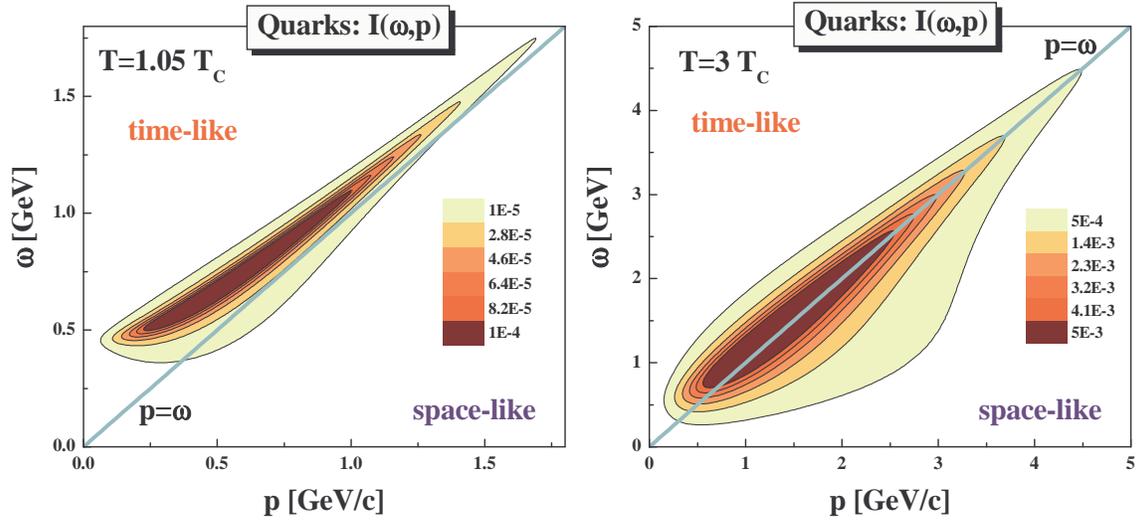}
    \caption{The integrand $I(\omega,p)$ (\ref{explain}) as a function
    of $\omega$ and $p$ for quarks at temperatures $T=1.05\  T_c$ (l.h.s.)
    and $T= 3\  T_c$ (r.h.s.). At $T=1.05\  T_c$ (l.h.s.) the
    quasiparticle mass amounts to $m \approx 0.55$ GeV and the width to
    $\gamma_q \approx$ 0.034  GeV while at $T=3\  T_c$ (r.h.s.) $m \approx$
    0.7 GeV and $\gamma_q \approx $ 0.25 GeV. The contour lines in both
    figures extend over one order of magnitude.  Note that for a convergence
    of the integrals (\ref{eq: N+}) the upper limits for
    $\omega$ and $p$ have to be increased by
    roughly an order of magnitude compared to the area shown in the figure.}
    \label{fig3}
\end{figure}

Before coming to the actual results for the quantities (\ref{eq:
N+}) and (\ref{scalar}) it is instructive to have a look at the
integrand in the quark density (\ref{eq: N+}) which reads as (in
spherical momentum coordinates with angular degrees of freedom
integrated out)
\be
\label{explain} I(\omega, p) =  \frac{d_q}{2 \pi^3}\ p^2 \ \omega
\, \rho_q(\omega,p^2)\, n_F((\omega-\mu_q)/T)  \, . \ee Here the
integration is to be taken over $\omega$ and $p$ from $0$ to
$\infty$. The integrand $I(\omega, p)$ is shown in Fig. \ref{fig3}
for $T=1.05 T_c$ (l.h.s.) and $T=3 T_c$ (r.h.s.) ($\mu_q=0$) in
terms of contour lines spanning both one order of magnitude. For
the lower temperature the quark mass is about 0.55 GeV and the
width $\gamma \approx $ 0.034 GeV such that the quasiparticle
properties are close to an on-shell particle. In this case the
integrand $I(\omega,p)$ is essentially located in the time-like
sector and the integral over the space-like sector is almost
negligible. This situation changes for $T = 3 T_c$ where the mass
is about 0.7 GeV while the width increases to $\gamma \approx $
0.25 GeV. As one observes from the r.h.s. of Fig. \ref{fig3} the
maximum of the integrand is shifted towards the line $\omega = p$
and higher momentum due to the increase in temperature by about a
factor of three; furthermore, the distribution reaches far out in
the space-like sector due to the Fermi factor $n_F(\omega/T)$
which favors small $\omega$. Thus the relative importance of the
time-like (+) part to the space-like (-) part is dominantly
controlled by the width $\gamma$ - relative to the pole mass -
which determines the fraction of $N_q^-$ with negative invariant
mass squared $(P^2 < 0)$ relative to the time-like part $N_q^+$
($P^2 > 0$).

\begin{figure}[htb!]
  \vspace{0.6cm}
    \includegraphics[width=11.5cm]{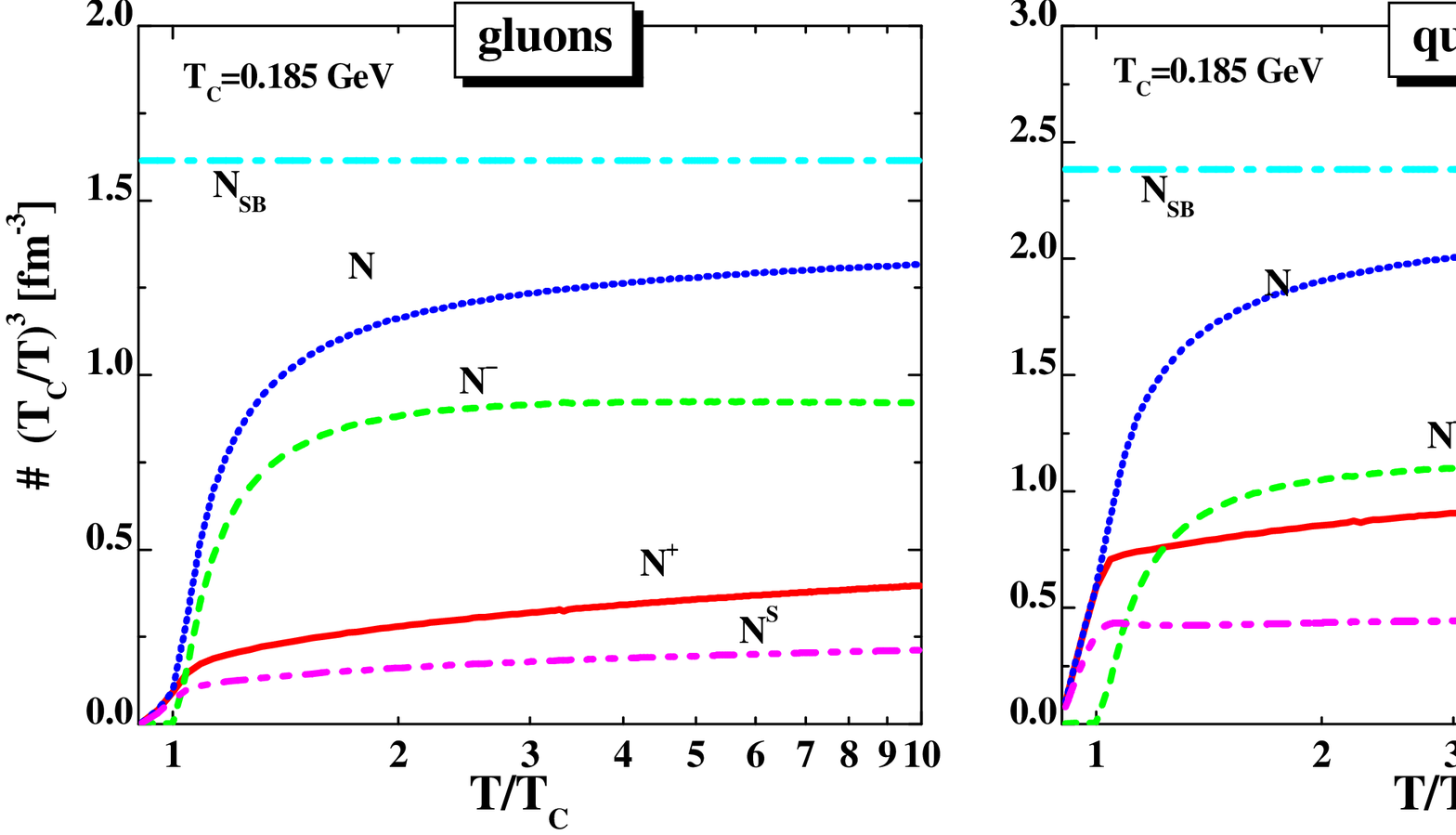}
    \caption{The various 'densities' (\ref{eq: N+}) for gluons (l.h.s.) and quarks (r.h.s.)
    including antiquarks (at $\mu_q=0$). The lower magenta
lines represent the scalar densities $N^s$, the red solid lines
the time-like densities $N^+$, the green lines the quantities
$N^-$ while the thick solid blue lines are the sum $N=N^+ + N^-$
as a function of $T/T_c$.  The upper dash-dotted lines display the
Stefan Boltzmann limits $N_{SB}$ for reference. All densities are
multiplied by the dimensionless factor $(T_c/T)^3$ to divide out
the leading scaling with temperature. }
    \label{fig4}
\end{figure}

The actual results for the different 'densities' (multiplied by
$(T_c/T)^3$) are displayed in  Fig. \ref{fig4} for gluons (l.h.s.)
and quarks (r.h.s.) including the antiquarks. The lower (magenta)
lines represent the scalar densities $N^s$, the red solid lines
the time-like densities $N^+$, the green lines the quantities
$N^-$ while the thick solid blue lines are the sum $N=N^+ + N^-$
as a function of $T/T_c$ (assuming  $T_c$ = 0.185 GeV). It is
seen that $N^+$ is substantially smaller than $N^-$ in case of
gluons in the temperature range 1.1 $\leq T/T_c \leq$ 10. Except
for an overall scale factor the result for gluons is practically
the same as in the pure Yang-Mills case (cf. Fig. 3 in Ref.
\cite{Cassing06}). The quantity $N$ follows closely the Stefan
Boltzmann limit $N_{SB}$ for a massless noninteracting system of
bosons which is given in Fig. \ref{fig4} (l.h.s.) by the upper
dash-dotted line. Though  $N$ differs by less than 20\% from the
Stefan Boltzmann (SB) limit for $T > 2 T_C$ the physical
interpretation is essentially different! Whereas in the SB limit
all gluons move on the light cone without interactions only a
small fraction of gluons can be attributed to quasiparticles with
density $N^+$ within the DQPM that propagate within the lightcone.
The space-like part $N^-$ corresponds to 'gluons' exchanged in
$t$-channel scattering processes and thus cannot be propagated
explicitly in off-shell transport approaches without violating
causality and/or Lorentz invariance. In case of quarks (or
antiquarks) the results are qualitatively similar but now the
time-like part $N^+$ comes closer to the space-like part $N^-$
since the ratio of the width to the pole mass ($\gamma_q/m$) is
smaller than the corresponding ratio for gluons as stated above.
Furthermore, the quantity $N$ is closer to the respective SB limit
(for massless fermions) due to the lower effective mass of the
quarks. In this respect the quarks and antiquarks are closer to
(but still far from) the massless on-shell quasiparticle limit.

The scalar densities $N^s$ (lower magenta lines) follow smoothly the time-like densities
$N^+$ (for gluons as well as quarks+antiquarks) as a function of
temperature and uniquely relate to the corresponding time-like
densities $N^+$ or the temperature $T$ in thermal equilibrium.

The separation of $N^+$ and $N^-$ so far has no direct dynamical
implications except for the fact that only the fraction $N^+$ can
explicitly be propagated in transport models as argued above. Following
Ref. \cite{Cassing06} we, furthermore, consider the energy densities,
 \be \label{energy} T_{00,x}^\pm(T) = {\rm {\tilde Tr^\pm_x
}}\ \omega  \ , \ee
that specify time-like and space-like contributions
to the quasiparticle energy densities ($x= g, q , {\bar q}$).

\begin{figure}[htb!]
\vspace{0.5cm}
    \includegraphics[width=11.5cm]{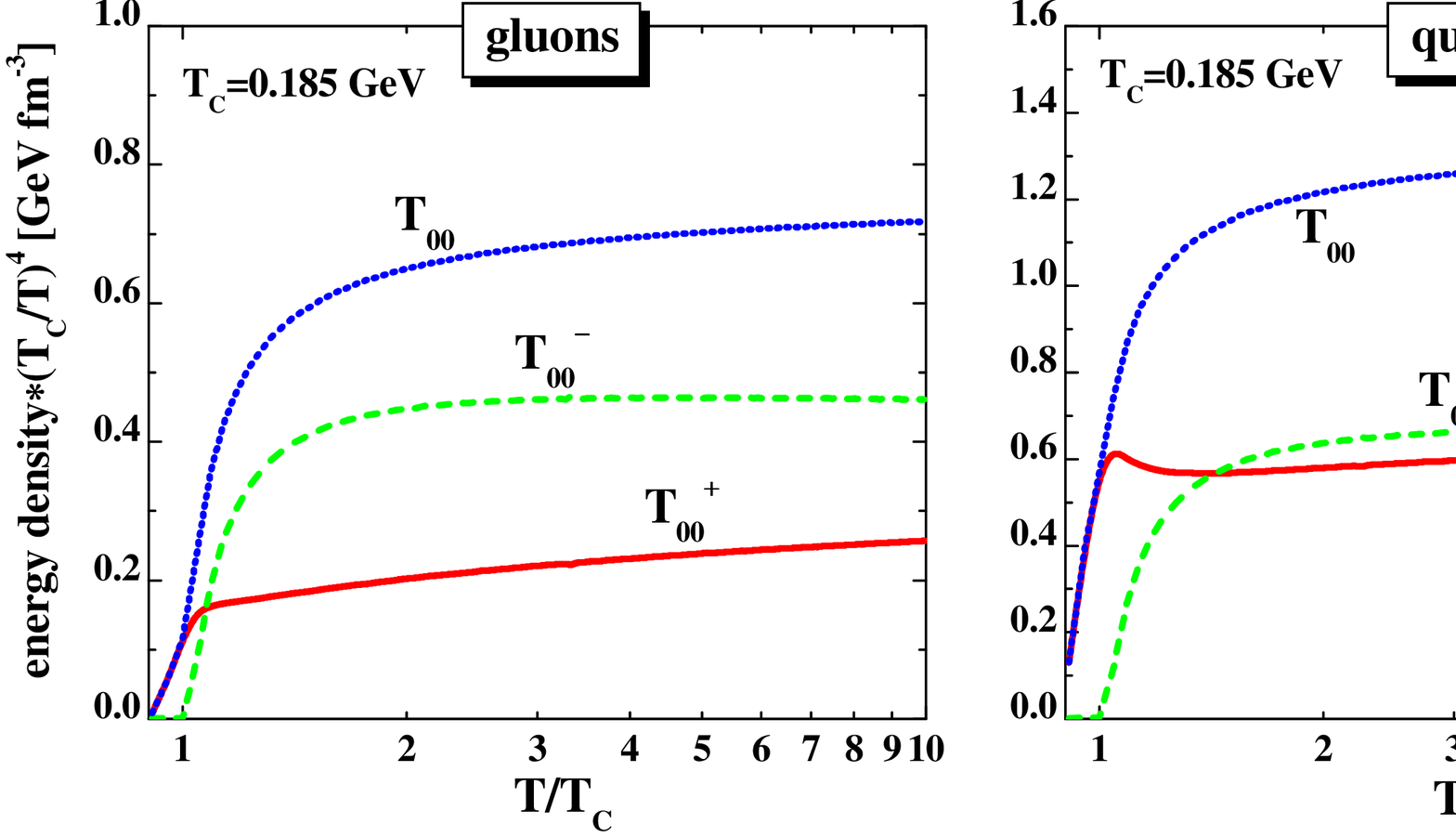}
    \caption{The time-like energy density $T_{00}^+$ (red solid line),
 the space-like energy density $T_{00}^-$  (dashed green line)
and their sum $T_{00}=T_{00}^+ + T_{00}^-$ (upper blue line)
as a function of $T/T_c$ for gluons (l.h.s.) and quarks (+ antiquarks)
(r.h.s.).  All densities are multiplied
by the dimensionless factor $(T_c/T)^4$. }
    \label{fig5}
\end{figure}

The result for the quasiparticle energy densities $T^+_{00}$ and
$T^-_{00}$ are displayed in Fig. \ref{fig5} for gluons (l.h.s.)
and quarks+antiquarks (r.h.s.) as a function of $T/T_c$. All
quantities have been multiplied by the dimensionless factor
$(T_c/T)^4$ in order to divide out the leading temperature
dependence. The lower red solid lines show the time-like
components $T_{00}^+$ while the dashed green lines display the
space-like parts $T_{00}^-$ which dominate over the time-like
parts except in the vicinity of $T_c$. The general behaviour for
the scaled energy densities $T_{00}^\pm$ is similar as for the
'densities' $N^\pm$ given in Fig. \ref{fig4} since the extra
factor $\omega$ in the integrand does not change significantly the
time-like and space-like parts. As in Ref. \cite{Cassing06} the
space-like parts are interpreted as potential energy densities
while the time-like fractions are the gluon and quark
quasiparticle contributions which propagate within the lightcone.

Summing up the time-like and space-like contributions for gluons,
quarks and antiquarks we obtain the total energy density $T^{00}$,
\be \label{ent} T^{00} = T_{00,g}^+ + T_{00,g}^- + T_{00,q}^+ +
T_{00,q}^- + T_{00,{\bar q}}^+ + T_{00,{\bar q}}^- \ , \ee which
is displayed in Fig. 2 by the dashed red line (scaled by
$T_c/T)^4$). As in case of the pure Yang-Mills system in Ref.
\cite{Cassing06} the quantity $T^{00}$ practically coincides with
the energy density $\epsilon$ (\ref{eps}) obtained from the
thermodynamical relations. Small differences of less than 5\% show
up  which indicates that the DQPM in its present formulation is
not fully consistent in the thermodynamical sense. Since these
differences are small on an absolute scale and significantly
smaller than differences between present independent lQCD
calculations for 3 quark flavors one may consider $T^{00}(T)
\approx \epsilon(T)$ and separate the kinetic energy densities
$T^+_{00}$ from the potential energy densities $T^-_{00}$ as a
function of the temperature $T$ or - in equilibrium - as a
function of the scalar densities $N^s$ or time-like densities
$N^+$, respectively.

\begin{figure}[htb!]
\vspace{0.8cm}
    \includegraphics[width=11.5cm]{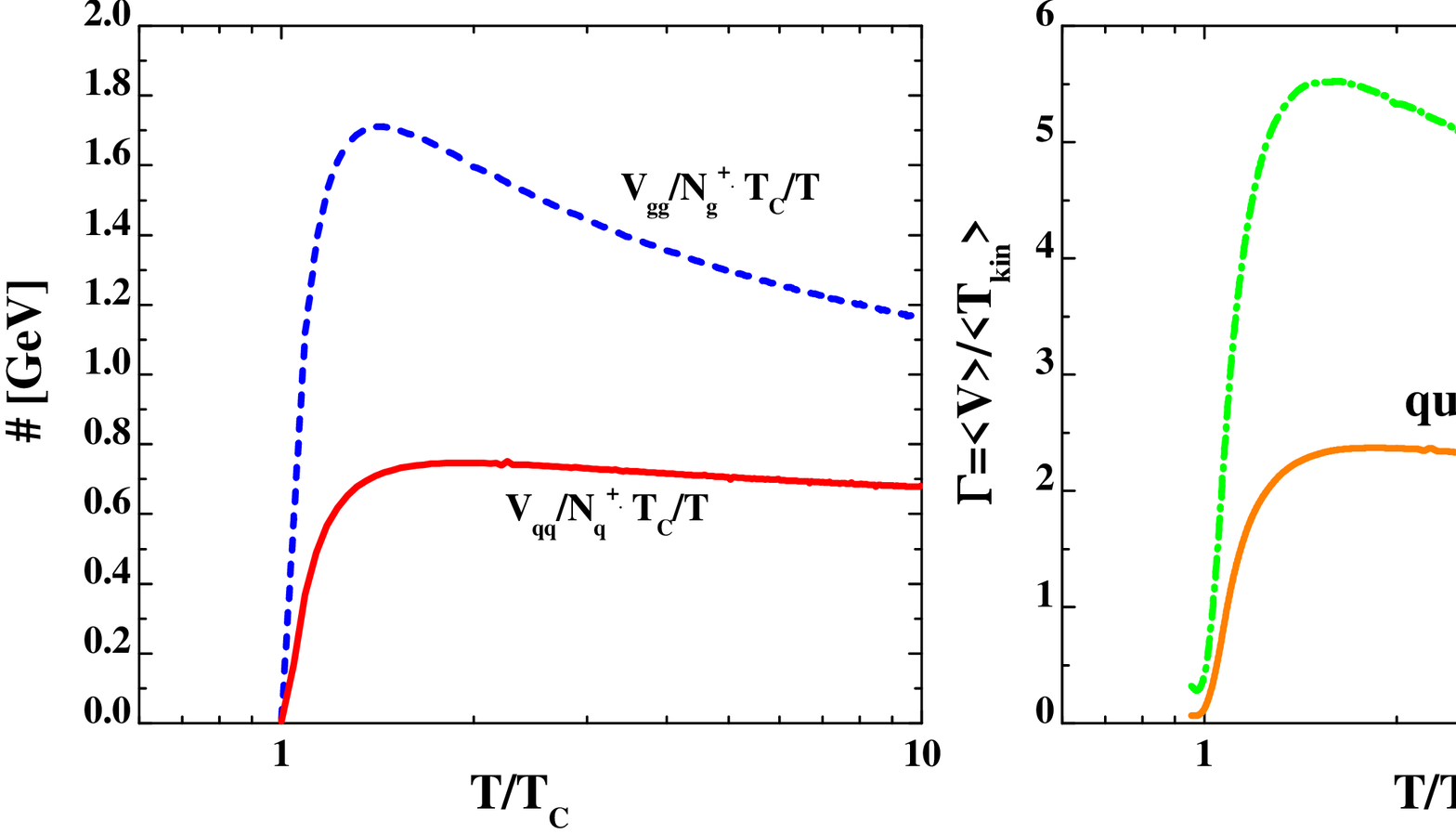}
    \caption{l.h.s.: The  potential energies per degree of
freedom  for gluons $V_{gg}/N^+_g = T_{00,g}^-/N^+_g$ (dashed blue
line) and for quarks (antiquarks) $V_{qq}/N^+_q =
T_{00,q}^-/N^+_q$ (solid red line) as a function of $T/T_c$. All
energies are multiplied by the dimensionless factor $(T_c/T)$.
r.h.s.: The plasma parameter $\Gamma$ (\ref{Plasma}) for gluons
(dot-dashed green line)  and quarks (solid orange line) as a
function of $T/T_c$. Note that $\Gamma \approx 1-2$ separates a
gas phase from a liquid phase in case of Lenard-Jones type of
interactions.}
    \label{fig6}
\end{figure}

It is instructive to show the  potential energies per degree of
freedom  $V_{gg}/N^+_g = T_{00,g}^-/N^+_g$ and $V_{qq}/N^+_q =
T_{00,q}^-/N^+_q$ as a function of  $T/T_c$. The corresponding
quantities are displayed in Fig. \ref{fig6} (l.h.s.) multiplied by
$T_c/T$ in terms of the solid red line and the dot-dashed blue
line.  It is seen that the potential energies per degree of
freedom steeply rise in the vicinity of $T_c$ and then increase
approximately linear with temperature $T$. As expected from the
larger width of the gluons the latter also show a potential energy
per degree of freedom which is roughly a factor of two larger than
the corresponding quantity for quarks (antiquarks). Consequently
rapid changes in the temperature (or density) - as in the
expansion of the fireball in ultrarelativistic nucleus-nucleus
collisions - are accompanied by a dramatic change in the potential
energy density and thus to a violent acceleration of the
quasiparticles. It is speculated here that the large collective
flow of practically all hadrons seen at RHIC \cite{STARS} might be
attributed to the early strong partonic forces expected from the
DQPM.

Furthermore, the plasma parameter $\Gamma$  defined by the ratio
of the average potential energy per particle to the average
kinetic energy per particle, \be \label{Plasma} \Gamma_g =
\frac{V_{gg}}{T_{kin,g}} \ , \hspace{2cm} \Gamma_q =
\frac{V_{qq}}{T_{kin,q}} \ , \ee is displayed in the r.h.s. of
Fig. \ref{fig6} for gluons (dot-dashed green line) and quarks
(solid orange line) as a function of $T/T_c$. Here the kinetic
energy densities are evaluated as \cite{Andre} \be \label{kinetic}
T_{kin,g} = {\rm {\tilde Tr}}^+_g (\omega - \sqrt{P^2}),
\hspace{2cm} T_{kin,q} = {\rm {\tilde Tr}}^+_q (\omega -
\sqrt{P^2}). \ee The presents results clearly indicate that the
plasma parameters $\Gamma_g, \Gamma_q$ are larger than unity
for both quarks and gluons
up to 10 $T_c$ (except for the vicinity of $T_c$) such that the
system should be in a liquid phase provided that some attractive
interaction between the constituents persists. Note that the
present evaluation of the plasma parameter $\Gamma$ is entirely
carried out within the DQPM and no longer based on estimates for
the potential energy as in Refs. \cite{Thoma,Andre}. Whereas in
the earlier estimates $\Gamma_g$ was dropping with temperature and
becoming lower than unity for $T > 4\  T_c$ \cite{Andre} the present
results indicate than the sQGP should persist for a much larger
range in temperature (or energy density) and thus also show up in
nucleus-nucleus collisions at Large Hadron Collider (LHC)
energies. Consequently, a partonic liquid is expected to be seen
also at LHC energies and as a consequence the observed scaling of
elliptic flow of practically all hadrons with the number of
constituent quarks (as seen at RHIC) should persist also at LHC.

\section{Selfenergies and effective interactions of time-like quasiparticles}
Since in transport dynamical approaches there are no
thermodynamical Lagrange parameters like the inverse temperature
$\beta = T^{-1}$ or the quark chemical potential $\mu_q$, which
have to be introduced in thermodynamics in order to specify the
average values of conserved quantities (or currents in the
relativistic sense), derivatives of physical quantities with
respect to the scalar densities $ N^s_x$  (or time-like
densities $ N^+_x$) ($x=g,q,{\bar q}$) are considered in the following (cf.
Refs. \cite{Cassing06,Toneev}).

\begin{figure}[htb!]
  \vspace{0.8cm}
    \includegraphics[width=11.cm]{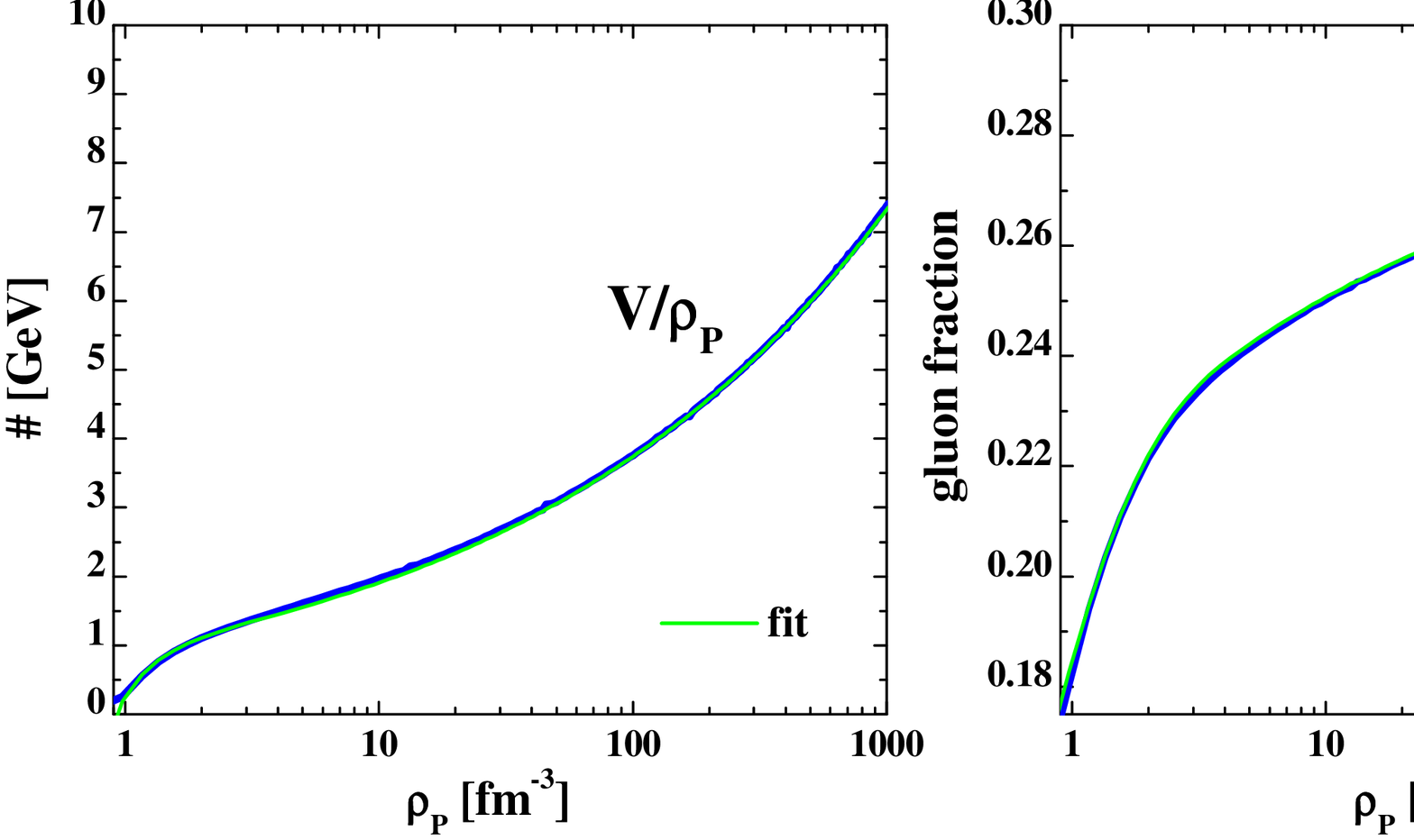}
    \caption{The parton
potential energy density $V$ (\ref{potp})  - divided by the parton
density $\rho_p$ -  from the DQPM (l.h.s., blue line)  as a
function of the parton density  $\rho_p$ (\ref{partond}). The
functional form of $V(\rho_p)$ is well reproduced by the
expression (\ref{fit1}) when divided by $\rho_p$ (green line) such
that the lines cannot be separated by eye. The r.h.s. shows the
gluon fraction $\alpha$ (\ref{partond}) from the DQPM (dark blue
line) as a function of $\rho_p$ together with the fit (\ref{fit2})
(green line). Again both lines cannot be distinguished by
eye.}    \label{fig7}
\end{figure}

The independent  potential energy densities  $V_x : = T_{00,x}^-$
now can be expressed as functions of the scalar densities $N^s_x$
(or $N^+_x$) instead of the temperature $T$ (and/or quark chemical
potential $\mu_q$). For a determination of mean-field potentials
for gluons and quarks (antiquarks) it is useful to consider the
partonic potential energy density \be \label{potp} V := T_{00,g}^-
+ T_{00,q}^- + T_{00,{\bar q}}^- = {\tilde V}_{gg} + {\tilde
V}_{qq} + {\tilde V}_{qg} \ee and to separate a pure gluonic
interaction density ${\tilde V}_{gg}$ from a pure fermionic
interaction density ${\tilde V}_{qq}$ as well as a gluon-fermion
interaction density ${\tilde V}_{qg}$. Correspondingly, a parton
density $\rho_p$ and gluon fraction $\alpha$ is defined via \be
\label{partond} \rho_p = N^+_g + N^+_q + N^+_{\bar q} \ ,
\hspace{2cm} \alpha = \frac{N^+_g}{N^+_g + N^+_q + N^+_{\bar q}} \
. \ee In the present DQPM (for $T_c$ = 0.185 GeV) the parton
density $\rho_p$ (\ref{partond}) turns out to be a simple function
of temperature, \be \label{pde} \rho_p(\frac{T}{T_c}) \approx
\left( \frac{T}{T_c} \right)^{3.15} \hspace{1.0cm} {\rm [fm^{-3}]
} \ , \ee such that the average distance between the partons is
given by $d(T/T_c) = \rho_p^{-1/3} \approx (T_c/T)^{1.05}$ [fm]
$\approx T_c/T$ [fm]. These relations allow to convert
temperatures scales to geometrical scales in a simple fashion.

In Fig. \ref{fig7} the parton potential energy density $V$
(\ref{potp}) - divided by the parton density $\rho_p$ -  is shown
as a function of $\rho_p$ (l.h.s.). The functional dependence of
$V$ on $\rho_p$ can be well approximated by the expression (cf.
l.h.s. of Fig. \ref{fig7}) \be \label{fit1} V(\rho_p) \approx
0.975 \rho_p^{1.292} - 0.71 \rho_p^{-2.1} \ \ [{\rm GeV/fm^3}] \ ,
\ee where the numbers in front carry a dimension in order to match
the units in GeV/fm$^3$. The gluon fraction $\alpha$
(\ref{partond}) is shown on the r.h.s. of Fig. \ref{fig7} and is
well approximated by \be \label{fit2} \alpha(\rho_p) = 0.29-0.075
\rho_p^{-0.28} -0.15 \exp(-1.6 \rho_p). \ee

Adding half of the interaction density ${\tilde V}_{qp}$ to the
gluon part and fermion part separately, we have $T_{00,g}^- =
{\tilde V}_{gg} + 0.5 {\tilde V}_{qg}$ and $T_{00,q}^- +
T_{00,{\bar q}}^- = {\tilde V}_{qq} + 0.5 {\tilde V}_{qg}$ such
that ${\tilde V}_{qq} - {\tilde V}_{gg} =: \Delta {\tilde V} =
T_{00,q}^- + T_{00,{\bar q}}^- - T_{00,g}^-$. The relative
fraction of this quantity to the total potential energy density
then is evaluated as \be \label{frac5} \kappa(\rho_p) =
\frac{\Delta {\tilde V}}{V} = \frac{\Delta {\tilde V}}{T_{00,q}^-
+ T_{00,{\bar q}}^- + T_{00,g}^-} . \ee Using the {\it Ansatz}:
\be \label{ansatz2} {\tilde V}_{gg}+{\tilde V}_{qq} = (1-\xi) V
\ee then gives \be  {\tilde V}_{gg} = 0.5 (1-\xi - \kappa) V \ ,
\hspace{1cm} {\tilde V}_{qq} = 0.5 (1-\xi + \kappa) V \ ,
\hspace{1cm}  {\tilde V}_{qg} = \xi V \ , \ee with still unknown
fraction $\xi$ for the interaction density ${\tilde V}_{qg}$.

In order to determine mean-field potentials $U_g (\rho_p)$ for
gluons or $U_q(\rho_p)$ for quarks (in the rest frame of the
system) one has to consider the derivatives (cf. Ref.
\cite{Cassing06,SIGMAM}) \be \label{mfields} U_g(\rho_p) : =
\frac{\partial ({\tilde V}_{gg} + {\tilde V}_{qg})}{\partial
N_g^+} \  , \hspace{1cm} U_q(\rho_p) : = \frac{\partial ({\tilde
V}_{qq} + {\tilde V}_{qg})}{\partial (N_q^+ + N_{\bar q}^+)} \ ,
\ee which by virtue of (\ref{ansatz2}) can be computed as \be
\label{mfields2} \hspace{-0.3cm} U_g(\rho_p)  = \frac{1}{2}
\frac{\partial (1-\kappa+\xi) V }{\partial \rho_p} \frac{\partial
\rho_p}{\partial N_g^+} \  , \hspace{0.4cm} U_q(\rho_p) =
\frac{1}{2} \frac{\partial (1+\kappa+\xi) V }{\partial \rho_p}
\frac{\partial \rho_p}{\partial (N_q^+ + N_{\bar q}^+)} \ . \ee
The fraction $\xi$ of the interaction density - in principle a
function of $\rho_p$ but here taken to be a constant -  now can be
fixed in comparison to the gluon mean-field from Ref.
\cite{Cassing06} where the pure Yang-Mills sector has be
investigated in the same way. This leads to $\xi \approx 0.3$ and
separates the total potential energy density $V$ into $\approx 26
\%$ for the gluon-gluon interaction part, 30\% for the quark-gluon
interaction part (including the antiquarks) and $\approx$ 44\% for
the fermionic interaction part.

\begin{figure}[htb!]
\vspace{0.5cm}
    \includegraphics[width=11.5cm]{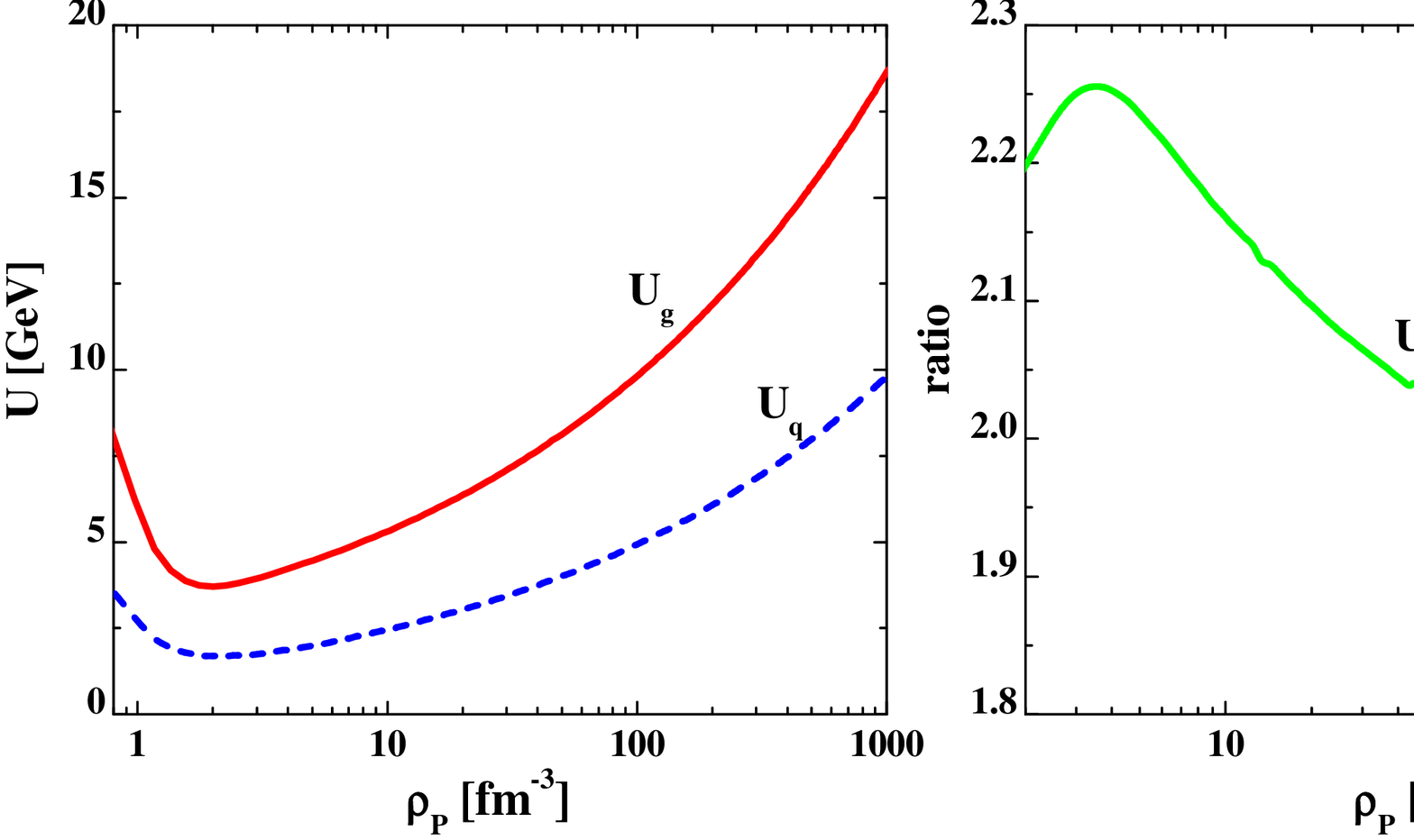}
    \caption{The mean-field potentials $U_g(\rho_p)$ for gluons
    (solid red line) and $U_q(\rho_p)$ for quarks (dashed blue
    line) as a function of the parton density $\rho_p$ (l.h.s.).
    The r.h.s. displays their ratio as a function of $\rho_p$ which is $\approx 2.05$
    within 10\% accuracy. }
    \label{fig8}
\end{figure}

The corresponding results for $U_g(\rho_p)$ and $U_q(\rho_p)$ are
displayed in the l.h.s. of Fig. \ref{fig8} in terms of the solid
red line and dashed blue line, respectively, and show distinct
minima at $\rho_p \approx$ 2.2 fm$^{-3}$ which corresponds to an
average partonic distance of $\approx$ 0.77 fm. The actual
numerical results for the mean-fields can be fitted by the
expressions,
\be
\label{pott} \hspace{1.7cm} U_g(\rho_p) \approx 70 \
e^{-\rho_p/0.31} + 2.65\ \rho_p^{0.21} + 0.45\  \rho_p^{0.4} \, \
\ [{\rm GeV}] \ , \ee $$ U_q(\rho_p) \approx 32 \ e^{-\rho_p/0.31}
+ 1.1\ \rho_p^{0.21} + 0.3\  \rho_p^{0.41} \, \ \ [{\rm GeV}] \ ,
$$

\noindent where $\rho_p$ is given in fm$^{-3}$ and the actual
numbers in front carry a dimension in order to match to the proper
units of GeV for the mean-fields. The ratio $U_g/U_q \approx 2.05$
as can be seen in the r.h.s. of Fig. \ref{fig8} for a very wide
range of parton densities $\rho_p$ within 10\% accuracy.

Some comments on the Lorentz structure of the mean fields $U_g$
and $U_q$ appear appropriate. Note that by taking the derivatives
with respect to the time-like densities one implicitly assumes
that a 4-vector current is the physical source of the selfenergies
and that $U_g, U_q$ are the 0'th components of  vector fields
$U^\nu_g$ and $U^\nu_q$ ($\nu = 0,1,2,3$). The spatial components
are assumed to vanish in the rest frame of the system and can be
evaluated by a proper Lorentz boost to the frame of interest. This
implies that the dynamical forces (as space-time derivatives of a
Lorentz vector) are Lorentz tensors as in case of QED (or vector
selfenergies as in the nuclear physics context \cite{SIGMAM}). On
the other hand one might consider the notion of purely scalar
selfenergies where derivatives of the potential energy density
with respect to the scalar density (e.g. $\partial
T_{00,x}^-/\partial N^s_x$) define an effective mass $M_x^*$
\cite{SIGMAM}. In a selfconsistent framework then the
quasiparticles masses (\ref{eq:M2b}) should be given by the
derivatives with respect to the scalar densities. However, this
relation is not fulfilled at all since a numerical evaluation of
the scalar derivatives gives effective masses that are larger by
more than an order of magnitude as the quasiparticle masses
introduced in (\ref{eq:M2b})! This result might have been
anticipated since the effective forces for a gauge (vector) field
theory should be dominated by Lorentz forces as in case of QED.
For completeness we give the fitted expressions for the
quasiparticle masses (\ref{eq:M2b}) as a function of the scalar
densities, respectively, \be \hspace{2cm} M_g(N^s_g) \approx 0.7
(N^s_g)^{0.248} + 0.102 (N^s_g)^{-2/3} \hspace{1cm} {\rm [GeV]} \
, \ee $$ m_q(N^s_{q+{\bar q}}) \approx 0.36 (N^s_{q+{\bar
q}})^{0.257} + 0.14 (N^s_{q+{\bar q}})^{-2/3} \hspace{1cm} {\rm
[GeV]}, $$

\noindent where the scalar densities are given in units of
fm$^{-3}$ and the numbers in front carry a dimension to match the
units of GeV for the masses (using $N^s_{q+{\bar q}} =  N^s_q +
N^s_{\bar q}$).

 Some information on the
properties of the effective gluon-gluon, quark-gluon and
quark-quark interaction may be extracted from the second
derivatives of the potential energy density $V$, i.e. \be
\label{interaction}  \hspace{1cm} v_{gg}(\rho_p): = \frac{\partial^2 {\tilde
V}_{gg}}{\partial N_g^{+2}} \approx \frac{1}{2} \frac{\partial^2
(1-\xi-\kappa) V}{\partial \rho_p^2} \left( \frac{\partial
\rho_p}{\partial N_g^+} \right)^2\ , \ee $$ v_{qq}(\rho_p): =
\frac{\partial^2 {\tilde V}_{qq}}{\partial (N_q^{+}+N_{\bar
q}^+)^2}
 \approx
\frac{1}{2} \frac{\partial^2 (1-\xi+\kappa) V}{\partial \rho_p^2}
\left( \frac{\partial \rho_p}{\partial( N_q^+ + N_{\bar q}^+)}
\right)^2\ ,$$ $$ v_{qg}(\rho_p): = \frac{\partial^2 {\tilde
V}_{qg}}{\partial (N_q^{+}+N_{\bar q}^+) \partial N_g^+}
 \approx
 \frac{\partial^2 (\xi V)}{\partial
\rho_p^2} \left( \frac{\partial \rho_p}{\partial( N_q^+ + N_{\bar
q}^+)} \right)   \left( \frac{\partial \rho_p}{\partial N_g^+}
\right)  \ .$$

\begin{figure}[htb!]
\vspace{0.5cm}    \includegraphics[width=11.5cm]{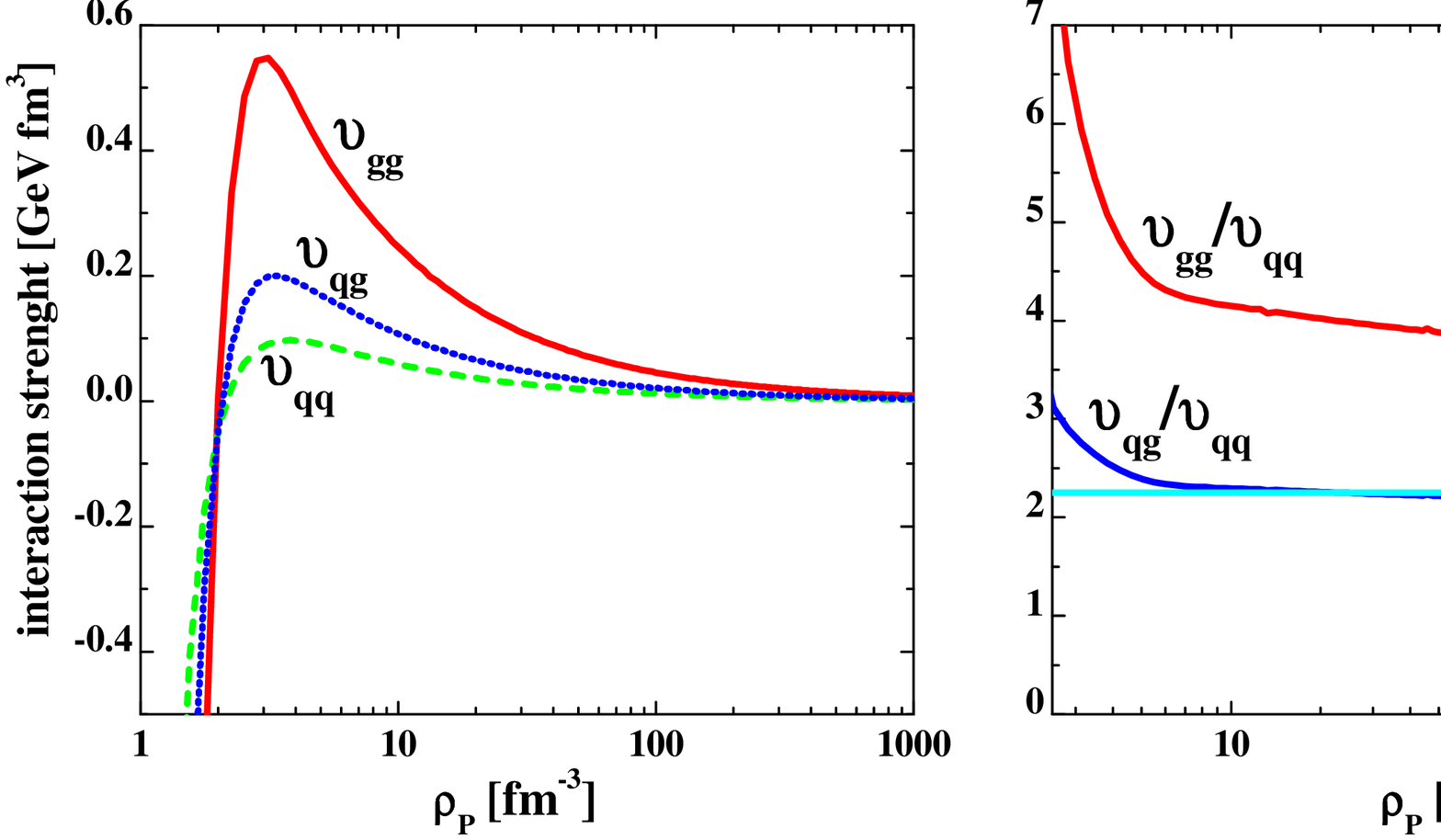}
    \caption{l.h.s. The effective gluon-gluon (solid red line),
    gluon-quark (solid blue line)
and quark-quark interaction (dashed green line) from the DQPM for
$\xi$ = 0.3 (see text). The r.h.s. displays the ratios
$v_{gg}/v_{qq}$ (solid red line) and $v_{qg}/v_{qq}$ (solid blue
line). The straight light blue line is the ratio of the Casimir
eigenvalues, i.e. $C_g/C_q$= 9/4.}
    \label{fig10}
\end{figure}

\noindent The numerical results for the interactions
(\ref{interaction}) are displayed in Fig. \ref{fig10} (l.h.s.) for
the effective gluon-gluon (solid red line), gluon-quark (solid
blue line) and quark-quark interaction (dashed green line). All
interactions show up to become strongly attractive at low parton
density $\rho_p < $ 2.2 fm$^{-3}$, change sign and become
repulsive for all higher parton densities. This situation has been
the same in the pure Yang-Mills case \cite{Cassing06} (at slightly
lower gluon density $\approx$ 1.4 fm$^{-3}$).
 Note that the
change of quasiparticle momenta (apart from collisions) will be
essentially driven by the (negative) space-derivatives $-\nabla
U_j(x) = - d U_j(\rho_p)/d \rho_p \ \nabla \rho_p(x)$ which
implies that the partonic quasiparticles (at low parton density)
will bind with decreasing density, i.e. form 'glueballs', mesons,
baryons or antibaryons dynamically close to the phase boundary and
repel each other for $\rho_p >$ 2.2 fm$^{-3}$. Note that color
neutrality is imposed by color-current conservation and only acts
as a boundary condition for the quantum numbers of the
bound/resonant states in color space.

The r.h.s. of Fig. \ref{fig10} displays the ratios $v_{gg}/v_{qq}$
(solid red line) and $v_{qg}/v_{qq}$ (solid blue line) as a
function of the parton density $\rho_p$ and demonstrates that
$v_{qg}/v_{qq} \approx 9/4$, which is the ratio of the Casimir
eigenvalues.

A straight forward way to model the parton condensation or
clustering to confined glueballs or hadrons dynamically (close to
the phase transition) is to adopt screened Coulomb-like potentials
$v_c(r,\Lambda)$ with the strength $\int d^3r \ v_c(r,\Lambda)$
fixed by the interactions $v_{gg}(\rho_p), v_{qg}(\rho_p),
v_{qq}(\rho_p)$ from (\ref{interaction}) and a screening length
$\Lambda$ from lQCD studies. For the 'dilute parton regime'
($\rho_p < $ 2.2 fm$^{-3}$), where two-body interactions should
dominate, one may solve a Schr\"odinger (Dirac or Klein-Gordon)
equation for the bound and/or resonant states. This task is not
addressed further in the present study since for the actual
applications in the PHSD approach \cite{PHSD} the formation of
glueballs is discarded and the formation of resonant hadronic
states close to the phase boundary is described by
density-dependent transition matrix elements
 ($\sim |v_{qq}(\rho_p)|^2$)
between partons of 'opposite' color with fixed flavor content. In
this way the energy-momentum conservation, the flavor current
conservation as well as 'color neutrality' are explicitly
fulfilled in the (PHSD) transport calculations for interacting
particles with spectral functions of finite width.

\section{Finite quark chemical potential $\mu_q$}
The extension of the DQPM to finite quark chemical potential
$\mu_q$ is more delicate since a guidance by lQCD is presently
very limited. In the simple quasiparticle model one may use the
stationarity of the thermodynamic potential with respect to
self-energies and (by employing Maxwell relations) derive a
partial differential equation for the coupling $g^2(T,\mu_q)$
which may be solved with a suitable boundary condition for
$g^2(T,\mu_q=0)$ \cite{pQP}. Once $g^2(T,\mu_q)$ is known one can
evaluate the changes in the quasiparticle masses with respect to
$T$ and $\mu_q$, i.e. $\partial M_x^2/\partial \mu_q$ and
$\partial M_x^2/\partial T$ (for $x=g,q,\bar{q}$) and calculate
the change in the 'bag pressure' $\Delta B$ (cf. Refs.
\cite{pQP,Reb} for details). However, such a strategy cannot be
taken over directly since additionally the quasiparticle widths
$\gamma_x(T,\mu_q)$ have to be known in the $(T,\mu_q)$ plane in
case of the DQPM.

In hard-thermal-loop (HTL) approaches \cite{BraP2,BlaJ} the
damping of a hard quark (or gluon) does not depend on the quark
chemical potential explicitly \cite{Vija} and one might  employ
(\ref{eq:gamma}) also at finite $\mu$. This, however, has to  be
considered with care since  HTL approaches assume small couplings
$g^2$ and should be applied at sufficiently high temperature,
only. Present lQCD calculations suggest that the
ratio of pressure to energy density, $P/\epsilon$, is approximately
independent on $\mu_q$ as a function of the energy density $\epsilon$
 \cite{Fodorx}. Accordingly, the functional dependence of the
quasiparticle width $\gamma$ on $\mu_q$ and $T$ has to be modeled
in line with 'lattice phenomenology' (see below).

Assuming three light flavors ($q= u,d,s)$ and all
chemical potentials to be equal ($\mu_u = \mu_d = \mu_s = \mu$)
equations (\ref{eq:M2}) and (\ref{eq:M2b}) demonstrate that the
effective gluon and quark masses are  a function of \be
\label{Tstar} T^{*2} = T^2+\frac{\mu^2}{\pi^2} . \ee  Since the
coupling (squared) (\ref{eq:g2}) is a function of $T/T_c$ a
straight forward extension of the DQPM to finite $\mu$ is to
consider the coupling as a function of $T^*/T_c(\mu)$ with  a
$\mu$-dependent critical temperature,   \be \label{Tstar2}
T_c(\mu) \approx T_c(\mu=0)(1 - \frac{1}{2\pi^2}
\frac{\mu^2}{T_c(0)^2}) \approx T_c(0)(1 - 0.05
\frac{\mu^2}{T_c(0)^2}) . \ee The coefficient in front of the
$\mu^2$-dependent part can be compared to lQCD calculations at
finite (but small) $\mu$ which gives $0.07(3)$ \cite{Karsch9}
instead of 0.05 in (\ref{Tstar2}). Consequently one has to expect
an approximate scaling of the DQPM results if the partonic width
is assumed to have the form (\ref{eq:gamma}),
\be  \label{gammamu} \hspace{2cm}
\gamma_g(T,\mu)   =
  N_c\  \frac{g^2(T^*/T_c(\mu))}{4 \pi} \, T \
  \ln\frac{2c}{g^2(T^*/T_c(\mu))} \, , \ee
  $$     \gamma_q(T,\mu)
  =
  \frac{N_c^2-1}{2 N_c} \frac{g^2(T^*/T_c(\mu))}{4 \pi} \,  T \
   \ln\frac{2c}{g^2(T^*/T_c(\mu))}
  \ ,
$$

\noindent where $g^2(T/T_c)$ has been replaced by $g^2(T^*/T_c(\mu))$.
 In fact, as will be demonstrated below, this choice leads to an approximate
independence of the potential energies per degree of freedom as a
function of $\mu_q$.  Nevertheless, the conjecture (\ref{gammamu})
should be explicitly controlled by lQCD studies for $N_f$=3 at
finite quark chemical potential. Unfortunately, this task is
presently out of reach and one has to live with the uncertainty in
(\ref{gammamu}) which is assumed in the following investigations.

\begin{figure}[htb!]
\vspace{0.5cm}
    \includegraphics[width=11.5cm]{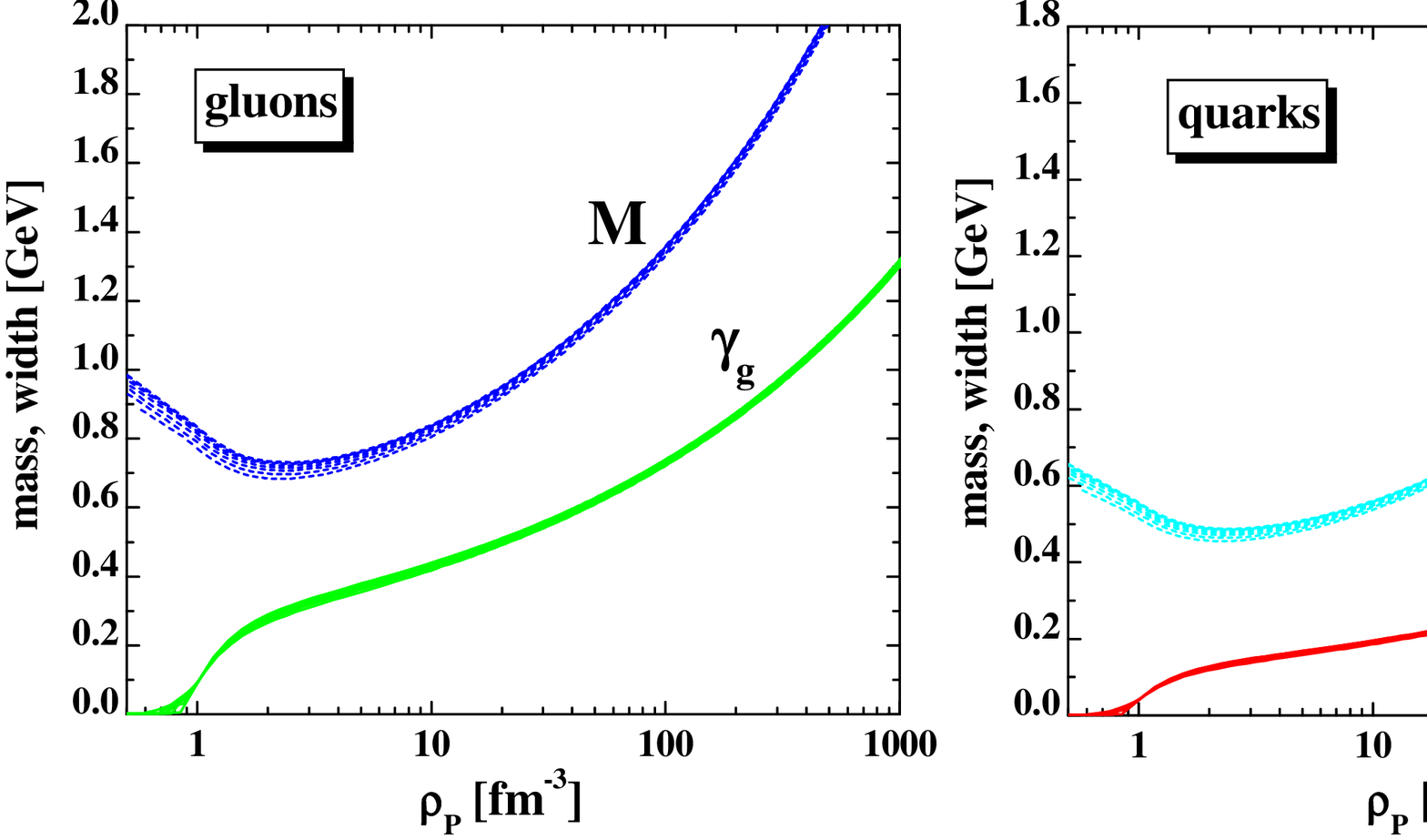}
    \caption{The gluon mass $M$ and width $\gamma_g$ (l.h.s.) as a function of the parton
    density $\rho_p$ for various chemical potentials $\mu_q = \mu$ from $\mu =0$ to $\mu=$ 0.21 GeV
    in steps of 0.03 GeV. The r.h.s. displays the mass $m$ for quarks and width $\gamma_q$
    for the same quark chemical potentials. }
    \label{fig18}
\end{figure}

Within the scaling hypothesis (\ref{Tstar2}), (\ref{gammamu}) the
results for the masses and widths in Section 2.1 stay
about the same as a function of $T^*/T_c(\mu)$ when dividing by
the temperature $T$. This also holds approximately when
displaying the masses and widths as a function of the parton
density $\rho_p$ for different chemical potentials $\mu$ as
demonstrated in Fig. \ref{fig18}. The latter quantities can well
be fitted by the expressions \be \label{fitmass} \hspace{3cm}
M_g(\rho_p) \approx 0.41 \ \rho_p^{0.255} + 0.38 \ \rho_p^{-0.7}
\hspace{1cm} {\rm [GeV]} \ , \ee  $$ \gamma_g(\rho_p) \approx 0.235
\ \rho_p^{0.245} - 0.14 \ \rho_p^{-2} \hspace{1cm} {\rm [GeV]} ,
$$ $$ m_q(\rho_p) \approx \frac{2}{3} M_g(\rho_p) , \hspace{1cm}
\gamma_q(\rho_p) \approx \frac{4}{9} \ \gamma_g(\rho_p)
 , $$

\noindent with $\rho_p$ given in units of fm$^{-3}$. Note that
according to the parametrization (\ref{fitmass}) the width might become negative for
very small $\rho_p$; in actual transport applications it should be set to
zero in such cases \cite{PHSD}.

\begin{figure}[htb!]
\vspace{0.5cm}
    \includegraphics[width=11.5cm]{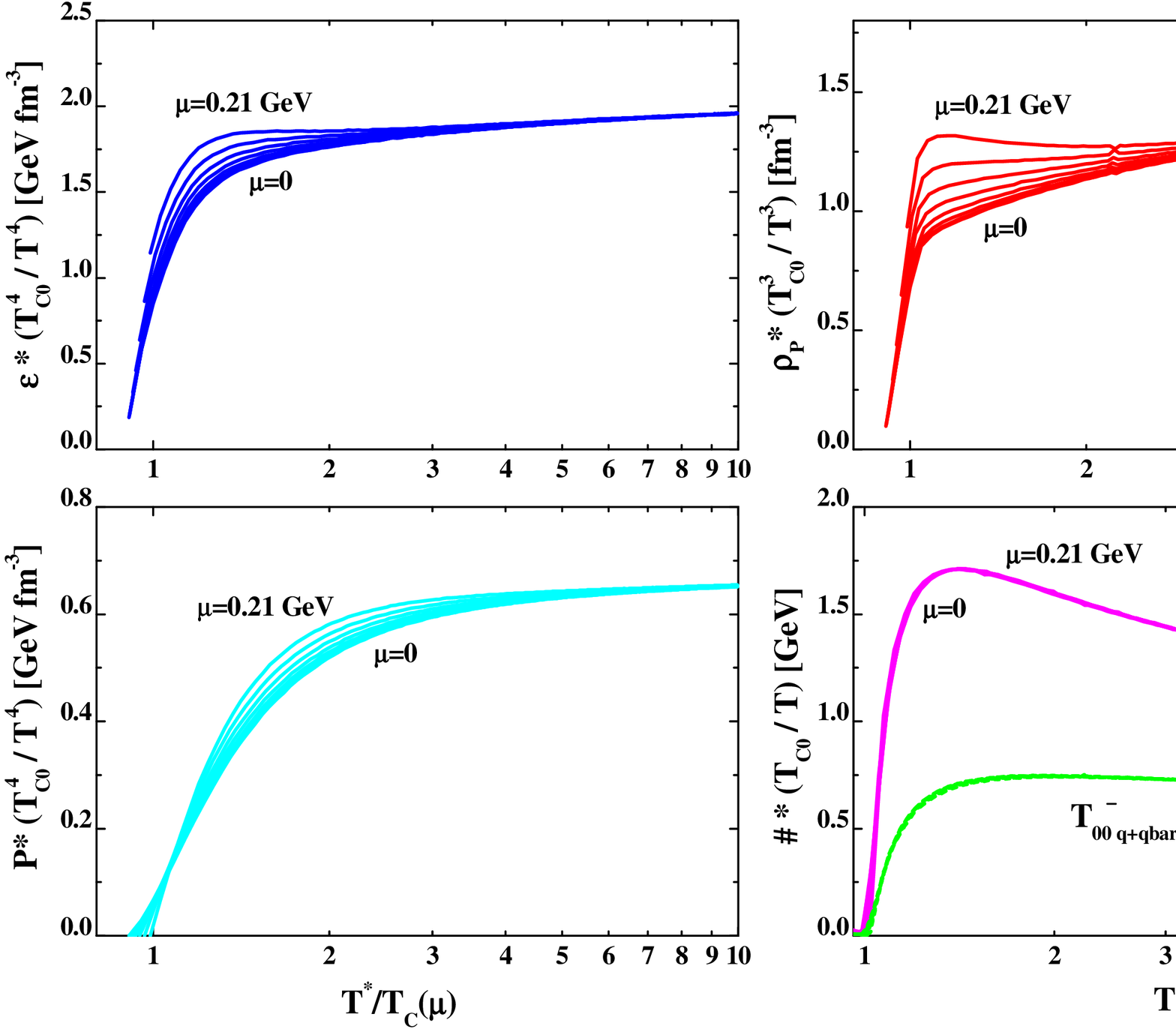}
    \caption{The energy density $\epsilon(T^*/T_c(\mu))$ (\ref{eps})
    for quark chemical potentials $\mu$ from 0 to 0.21 GeV in steps
    of 0.03 GeV as a function of the scaled temperature
    $T^*/T_c(\mu)$ for $N_f$ =3 (upper l.h.s.). The pressure $P(T^*/T_c(\mu))$
    from the thermodynamical relation (\ref{pressure})
    for quark chemical potentials $\mu$ from 0 to 0.21 GeV in steps
    of 0.03 GeV as a function of the scaled temperature
    $T^*/T_c(\mu)$ for $N_f$ =3 (lower l.h.s.). The parton density
    $\rho_p$ (for quark chemical potentials $\mu$ from 0 to 0.21 GeV) is shown in
    the upper r.h.s. The potential energy
    per time-like 'gluon' (upper magenta lines) and the potential
    energy per time-like 'quark+antiquark' (lower green lines) for
    the same     quark chemical potentials  are displayed  in the
    lower r.h.s. Note that $\epsilon$ and $P$ are
    scaled by the dimensionless factor $(T_{c0}/T)^4$ (with
    $T_{c0}$ = 0.185 GeV) while the parton density $\rho_p$ is scaled by the
    factor $(T_{c0}/T)^3$.}
    \label{fig14}
\end{figure}

 \noindent
The more interesting question is how the energy density $\epsilon$
(\ref{eps}) and the pressure $P$ (from (\ref{pressure})) change
with quark chemical potential $\mu=\mu_q$ in the DQPM. This
information is provided in Fig. \ref{fig14} where the upper l.h.s.
shows the energy density $\epsilon$ (\ref{eps}) (scaled in terms
of $T_{c0}=T_c(\mu=0)=$ 0.185 GeV ) as a function of $T^*/T_c(\mu)$.
Here a scaling of the 'temperature' $T^*$ with $T_c(\mu)$ (\ref{Tstar2})
is used since the phase boundary changes with the quark chemical
potential $\mu$. The energy density $\epsilon$ is seen to scale
well with $(T/T_{c0})^4$ as a function of temperature for
$T^*/T_c(\mu)
> 3$, however, increases slightly with $\mu$ close to the phase
boundary where the scaling is violated on the level of 20\%. This
violation in the scaling (seen in the upper left part of the
figure) is essentially due to an increase of the pressure $P$
which is displayed in the lower left part of the figure as a
function of $T^*/T_c(\mu)$ for the same chemical potentials $\mu$
from $\mu = 0$ to 0.21 GeV in steps of 0.03 GeV. Note that a quark
chemical potential of 0.21 GeV corresponds to a baryon chemical
potential of $\mu_B= 3 \mu = $0.63 GeV which is already substantial
and the validity of (\ref{Tstar2}) becomes questionable.

Since the pressure $P$ is obtained from an integration of the
entropy density $s$ over temperature (\ref{pressure}) the increase
in $P$ with $\mu$ can directly be traced back to a corresponding
increase in entropy density. The latter is dominated by the
time-like quasiparticle contributions thus 'counting' the
effective degrees of freedom, \be \label{rhog} \rho_p = N_g^+ +
N^+_{q+{\bar q}} \ , \hspace{2cm} N^+_{q+{\bar q}} = N_q^+ +
N_{\bar q}^+ \ . \ee

The upper r.h.s. of Fig. \ref{fig14} shows $\rho_p$ versus
$T^*/T_c(\mu)$ (multiplied by $T_{c0}^3/T^3$) for chemical
potentials from $\mu=0$ to $\mu$ = 0.21 GeV in steps of 0.03 GeV.
Indeed the scaled  parton density increases with $\mu$; this
increase is most pronounced for lower temperatures and becomes
substantial for $\mu$ = 0.21 GeV. One should recall, however, that
a tri-critical endpoint (in the $T,\mu$ plane) is expected for
$\mu_B = 3 \mu \approx$ 0.4 GeV \cite{FodorKatz} which corresponds
to $\mu \approx$  0.13 GeV. When restricting to the interval $0
\leq \mu \leq $ 0.13 GeV the explicit change in the parton density
$\rho_p$ with $\mu$ stays very moderate. This also holds for the
energy density $\epsilon$ and the pressure $P$.

Quite remarkably the potential energy per time-like fermion
$T^-_{00, q+{\bar q}}/N^+_{q+{\bar q}}$ changes very little with
$\mu$ as can be seen from the lower right part of Fig. \ref{fig14}
where the latter quantity is displayed for the same chemical
potentials as before (lower green lines) as a function of
$T^*/T_c(\mu)$. This also holds for the potential energy per time-like
gluon $T^-_{00,g}/N^+_g$ (upper magenta lines). Accordingly the potential
energy per time-like degree of  freedom is essentially a function of
$T^*/T_c(\mu)$ alone.

\begin{figure}[htb!]
\vspace{0.5cm}
    \includegraphics[width=11.5cm]{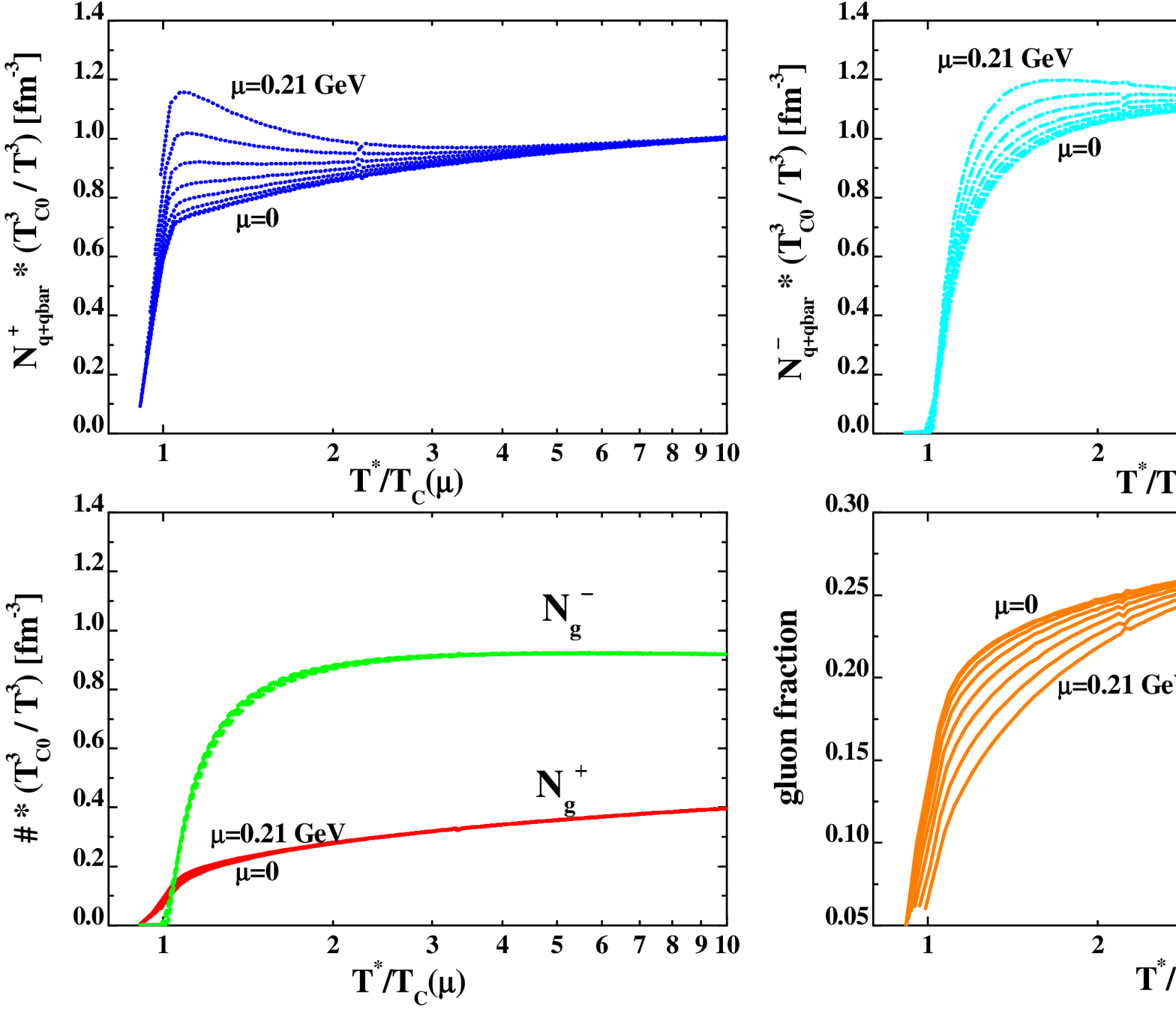}
    \caption{Upper parts: The time-like fermion density $N_{q+{\bar q}}^+$ (l.h.s., dark blue lines)
    and the space-like quantity $N^-_{q+{\bar q}}$ (r.h.s., light blue lines) for quark chemical
     potentials $\mu$ from 0 to 0.21 GeV in steps
    of 0.03 GeV as a function of the scaled temperature
    $T^*/T_c(\mu)$. Lower parts: the time-like gluon density
    $N_g^+$ (l.h.s., lower red lines) and the space-like quantity $N_g^-$
    (l.h.s., dashed green lines) for the same quark
    chemical potentials as a function of $T^*/T_c(\mu)$.
    Both, $N_g^-$ and $N_g^+$ change only very little with $\mu$
    in the whole temperature range; the relative changes are of
    the order of the line width.
    The gluon fraction (\ref{glufrac})
    is shown on the r.h.s. as a function of the scaled temperature  $T^*/T_c(\mu)$
    for quark chemical potentials $\mu$ from 0 to 0.21 GeV in steps
    of 0.03 GeV. Note that all 'densities' have been multiplied by the dimensionless
    factor $(T_{c0}/T)^3$.}
    \label{fig13}
\end{figure}

The time-like densities for the fermions $N^+_{q+{\bar q}}$
(\ref{rhog}) (upper l.h.s., dark blue lines) and the space-like
quantities $N^-_{q+{\bar q}}$ (upper r.h.s., light blue lines) are
shown in Fig. \ref{fig13} as a function of $T^*/T_c(\mu)$ for
chemical potentials $\mu$ from $\mu$ = 0 to $\mu=$ 0.21 GeV in
steps of 0.03 GeV (scaled by $T_{c0}^3/T^3$). Both quantities
increase with $\mu$ in a comparable fashion such that their ratio
stays approximately constant for $T^* > 1.5 \ T_c(\mu)$. The
space-like quantity $N_g^-$ (lower l.h.s., dashed green lines)
practically is independent from $\mu$ as well as the time-like gluon
density $N_g^+$ (lower l.h.s.,  red lines). Accordingly
the gluon fraction \be \label{glufrac} \alpha(T,\mu) =
\frac{N_g^+}{N^+_g + N^+_{q + {\bar q}}} \ ,\ee which is displayed on
the lower r.h.s. of Fig. \ref{fig13} for the same set of quark
chemical potentials as a function of $T^*/T_c(\mu)$, decreases with $\mu$.
It drops to zero below the phase boundary
because the difference between the gluon effective mass and the
fermion effective mass becomes large below $T_c$.

\begin{figure}[htb!]
\vspace{0.5cm}
    \includegraphics[width=11.5cm]{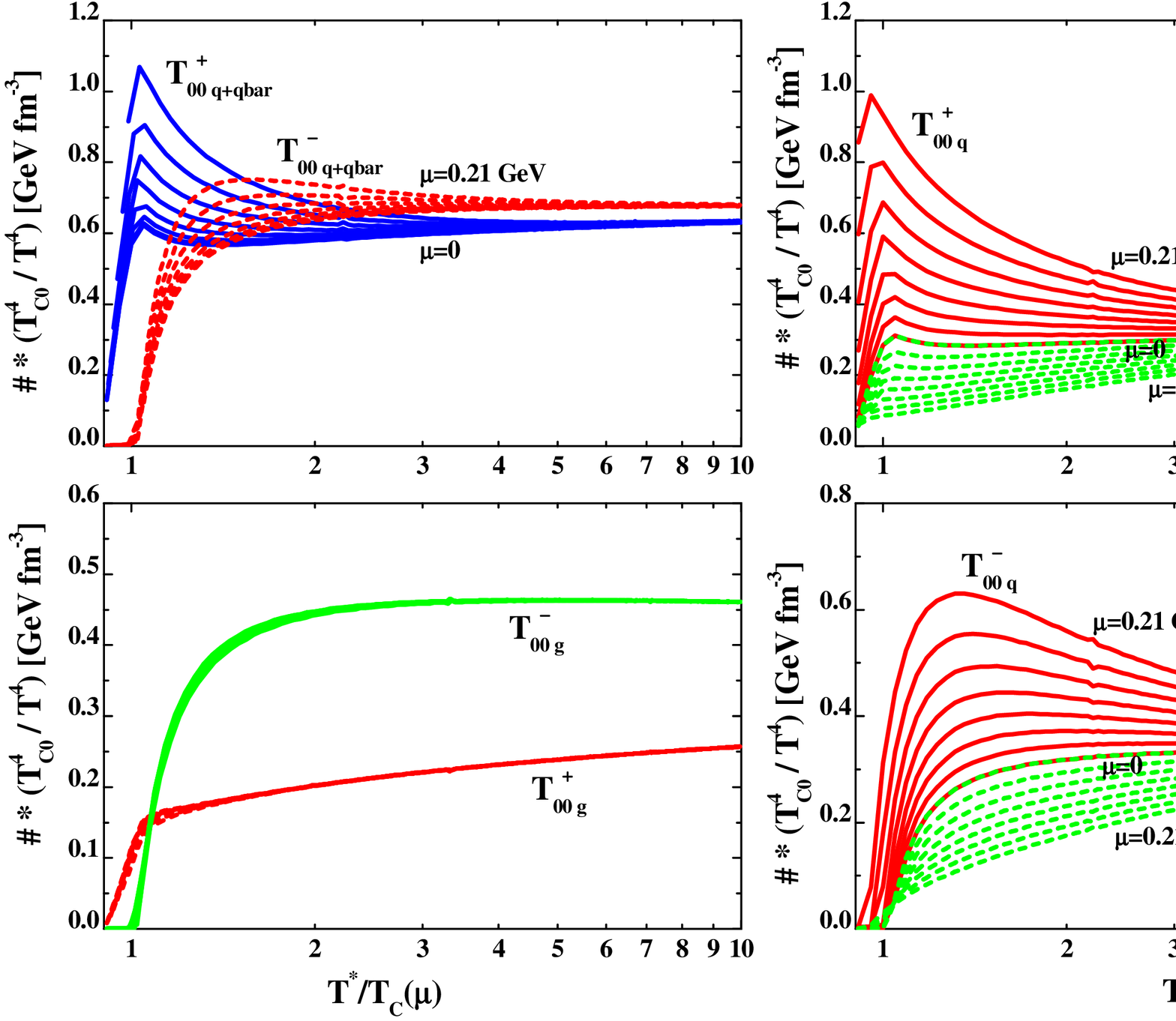}
    \caption{l.h.s.: The time-like fermion energy density $T^+_{00, q+{\bar q}}$ (blue lines)
    and the space-like  fermion energy density $T^-_{00, q+{\bar q}}$ (dashed red
    lines) for quark chemical
     potentials $\mu$ from 0 to 0.21 GeV in steps
    of 0.03 GeV as a function of the scaled temperature
    $T^*/T_c(\mu)$ (upper l.h.s.). The lower l.h.s. shows the
    time-like (dashed red lines) and space-like gluon energy
    density (green lines) for the same chemical potentials as a function of the
    scaled temperature. Both,  $T_{00, g}^-$ and  $T_{00, g}^+$ are roughly
    independent from $\mu$ in the whole temperature range.  The time-like (upper r.h.s) and
    space-like energy densities (lower r.h.s.) for quarks (upper
    red lines) and antiquarks (lower green lines) are separately displayed
     for quark chemical
     potentials $\mu$ from 0 to 0.21 GeV in steps
    of 0.03 GeV  for $N_f = 3$. All quantities have been multiplied by the dimensionless
    factor $T_{c0}^4/T^4$.}
    \label{fig15}
\end{figure}

We continue with the time-like and space-like energy densities for
the fermions and gluons as a function of $\mu$ and $T^*/T_c(\mu)$
which are displayed on the l.h.s. of Fig. \ref{fig15}. The upper
part shows that the space-like and time-like energy densities for
the fermions increase with $\mu$ roughly in a similar fashion as the fermion
'densities' such
that their ratio is approximately independent on $\mu$. The lower
part of Fig. \ref{fig15} (l.h.s.) indicates that the space-like
energy density for gluons (green lines) as well as the time-like
energy density for gluons (red lines)
are approximately independent from $\mu$ within line width.
 When separating the time-like fermion energy
density into contributions from quarks $q$ and antiquarks ${\bar
q}$ (r.h.s. of Fig. \ref{fig15}) we find an increase of $T_{00,
q}^+$ with $\mu$ which is not fully compensated by a decrease of
$T_{00, {\bar q}}^+$ with quark chemical potential. Similar
dependences on $\mu$ and $T^*$ are found for the space-like sector
(lower part, r.h.s.). Thus when summing up all space-like and
time-like energy densities from the fermions and the gluons a
small net increase in the total energy density with $\mu$
survives (see also upper left part of Fig. \ref{fig14}).

We note in passing that derivatives of the various energy
densities w.r.t. the time-like gluon or fermion densities (as
investigated in detail in Section 3) are approximately independent
of $\mu$ such that the effective potentials $U_g(\rho_p)$,
$U_q(\rho_p)$ and $U_q(\rho_p)$ (\ref{mfields2}) stay practically
the same. Since this result may be inferred already from the
$\mu$-(in)-dependence of the potential energy per time-like degree of freedom
(displayed in the lower r.h.s. of Fig. \ref{fig14}) an
explicit representation is discarded. This implies that the
mean-fields (\ref{mfields2}) or parametrizations (\ref{pott})  may
be employed also at finite (moderate) net quark density $N_q^+ -
N_{\bar q}^+$ which simplifies an implementation in parton
transport models (as e.g. PHSD).

\begin{figure}[htb!]
\vspace{0.5cm}
    \includegraphics[width=11.5cm]{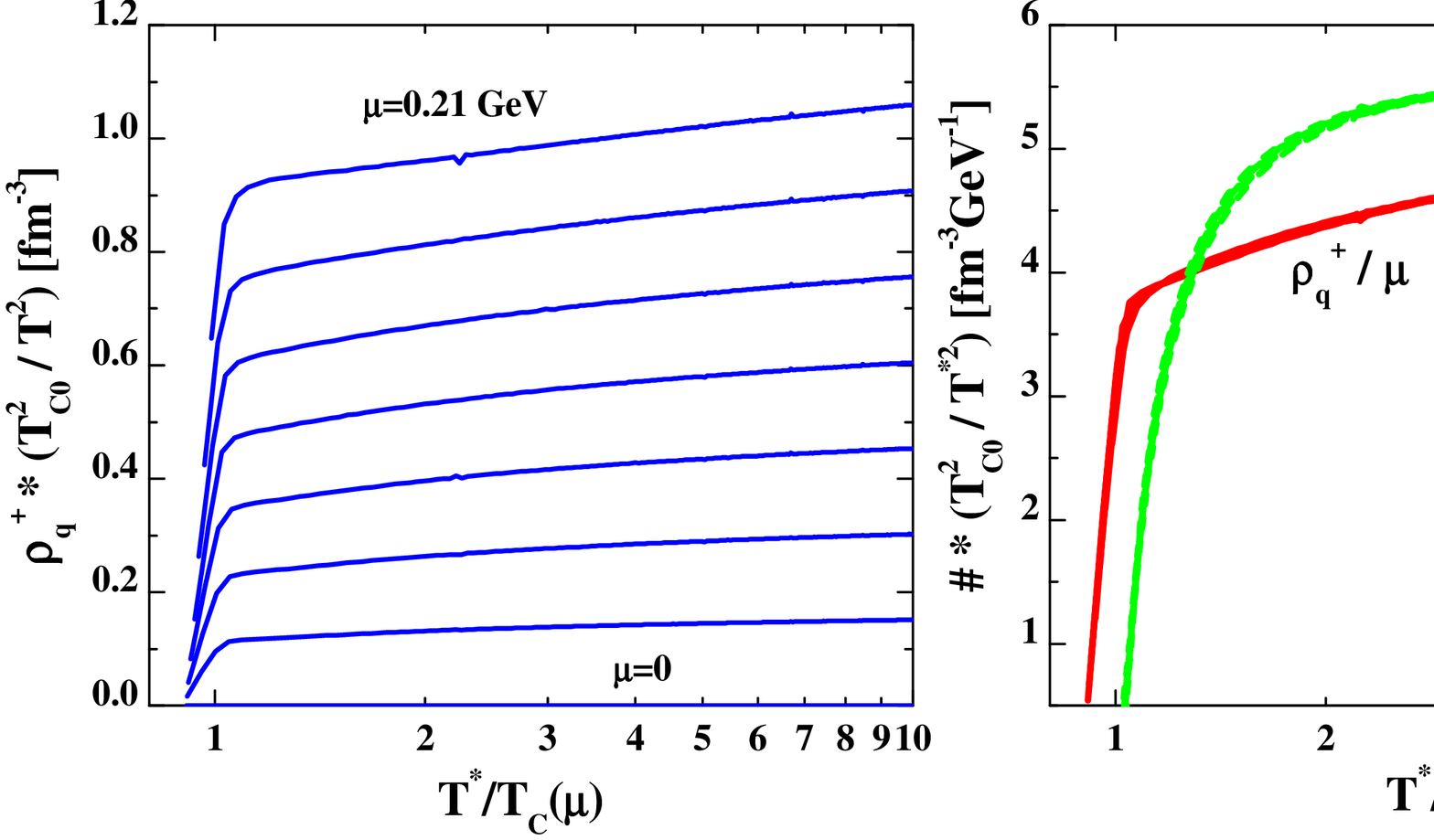}
    \caption{l.h.s.: The time-like quark density $\rho_q^+$ (\ref{qdens}) (scaled by
    the dimensionless factor $T^2_{c0}/T^2$)
    for quark chemical
     potentials $\mu$ from 0 to 0.21 GeV in steps
    of 0.03 GeV as a function of the scaled temperature
    $T^*/T_c(\mu)$. r.h.s.: The time-like quark density $\rho_q^+$ (\ref{qdens})
    (red solid lines) and the space-like quantity $\rho_q^-$
    (\ref{qdens})(dashed green lines)     for the same quark chemical
     potentials $\mu$. In this part of the figure all quantities are scaled by
    the factor $T^2_{c0}/T^{*2}/\mu$.  }
   \label{fig16}
\end{figure}

Whereas for vanishing quark chemical potential $\mu$=0 the quark
and antiquark densities are the same this no longer holds for
finite $\mu$ where the differences \be \label{qdens} \rho_q^{\pm}
= N_q^{\pm} - N_{\bar q}^{\pm} \ee are of separate interest since
$\rho_q = \rho_q^+ + \rho_q^-$ is the zero'th component of a
conserved flavor current (separately for each flavor $u,d,s,..$).
In order to obtain some idea about space-like and time-like
contributions of the net quark density $\rho_q$ in the DQPM we
first plot the time-like component $\rho_q^+$ in the l.h.s. of
Fig. \ref{fig16} as a function of $T^*/T_c(\mu)$ for different
$\mu$ from 0 to 0.3 GeV (as before). In order to divide out the
leading scaling with temperature  the time-like densities have
been multiplied by $T_{c0}^2/T^2$ on the l.h.s. of Fig.
\ref{fig16} (as known from Fermi systems in the nuclear physics
context). The actual results show an approximately linear increase
in $\mu$ (l.h.s.) which suggests to study the scaled quantities
$\rho_q^{\pm} \cdot T_{c0}^2/(\mu T^{2})$. However, a better
scaling is obtained when multiplying $\rho_q^{\pm}$ by
$T_{c0}^2/(\mu T^{*2})$. The latter quantities are displayed in
the r.h.s. of Fig. \ref{fig16} for the same set of chemical
potentials as before. This approximate scaling allows to estimate
the net quark density as \be \label{guess} \rho_q \approx 10 \ \mu
\ T^{*2}/T_{c0}^2 \ \ [{\rm fm}^{-3}] \ee
 with $\mu$ given in units of GeV in case of $N_f$ =3.

\begin{figure}[htb!]
  \begin{center}
\vspace{0.5cm}
    \includegraphics[width=9.5cm]{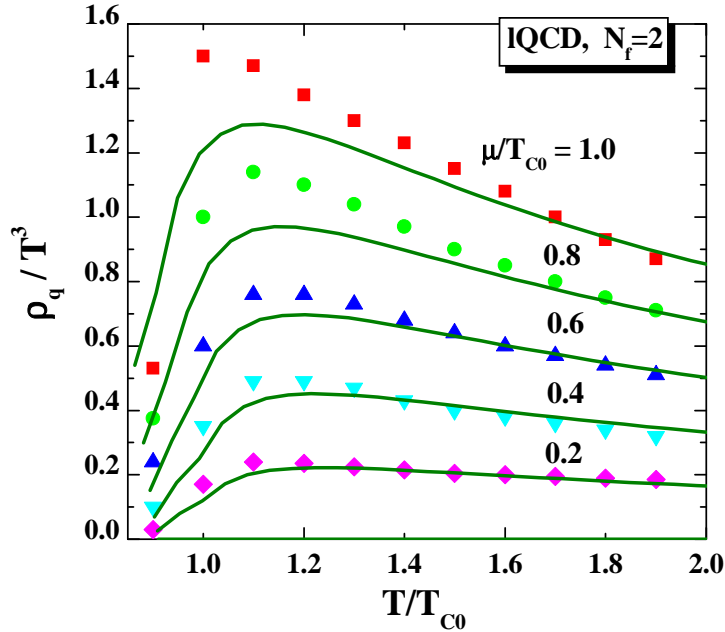}
    \caption{The quark density $\rho_q$ (\ref{qdens}) (divided by the factor
    $T^3$)    for quark chemical
     potentials $\mu/T_{c0}$ from 0 to 1.0 in steps of 0.2
     as a function of the scaled temperature
    $T/T_{c0}$ from the DQPM (green lines) for $N_f$ =2.
    The different symbols represent the lQCD results from Ref.
    \cite{Karsch8} for the same quark chemical potentials as a
    function of $T/T_{c0}$ (also for $N_f$ =2). Note that the temperature
    axis here is given by $T/T_{c0}= T/T_c(\mu=0)$ and not by $T^*/T_c(\mu)$!}
   \label{fig17}
  \end{center}
\end{figure}

The approximate scaling depicted in Fig. \ref{fig16} is a
prediction of the DQPM in case of 3 quark flavors and should be
controlled  by lQCD calculations. For two light flavors ($N_f =
2$) some comparison can be made in the low temperature (and low
$\mu$) range. Unfortunately the simple scaling relation
(\ref{guess}) does not hold at small $T$ such that an explicit
comparison has to be presented between the DQPM for $N_f$ = 2 and
the lQCD calculations from Ref. \cite{Karsch8}. The latter
calculations have been carried out on a 16$^3 \times 4$ lattice
with two continuum flavors (of p4-improved staggered fermions)
with mass $m= 0.4 T$. Though the fermion masses are not really
'light' in the lQCD calculations the actual lQCD results may serve
a test of the present DQPM in the 2-flavor sector (for low $\mu$
and $T$). The explicit lQCD results for the net quark density
$\rho_q$ (divided by $T^3$) are displayed in Fig. \ref{fig17} as a
function of $T/T_{c0}$ in terms of the various symbols. The
symbols of equal color correspond to $\mu/T_{c0}$ = 1.0, 0.8, 0.6,
0.4, and 0.2 from top to bottom and by eye show an approximately
linear dependence on $\mu$. The explicit results from the DQPM for
$N_f$ = 2 are presented in terms of the dark green lines and
approximately follow the lQCD results (at least for smaller
$\mu$). Since the systematic errors of the lQCD calculations are
not known to the author an explicit refitting of the parameters
$\lambda, T_s/T_c$ and $c$ for $N_f$ =2 (cf. Section 2.1) is
discarded here. Nevertheless, the qualitative (and partly
quantitative) agreement between the DQPM and lQCD provides a test
for the basic concepts of the DQPM.

\section{Dilepton radiation from the sQGP}
The properties of the sQGP so far have been fixed in the DQPM by
specifying the (vector) selfenergies/potentials as well as
effective interactions for the time-like partons. As shown in Ref.
\cite{Andre} this leads to a strongly interacting partonic system
with a shear viscosity to entropy density ratio close to $\eta/s \approx$
0.2. However, the predictions from the DQPM should be controlled
by independent lQCD studies to get some idea about the reliability
of the approach. As mentioned before transport coefficients like
the shear viscosity $\eta$ are available from lQCD \cite{lattice2}
but the present accuracy is not satisfactory. On the other hand
some lQCD information is available from the Bielefeld group on the
electromagnetic correlator which is initimately related to the
dilepton emission rate \cite{RappWa,Karsch6,Karsch7}.

\begin{figure}[htb!]
  \begin{center}
    \includegraphics[width=10.5cm]{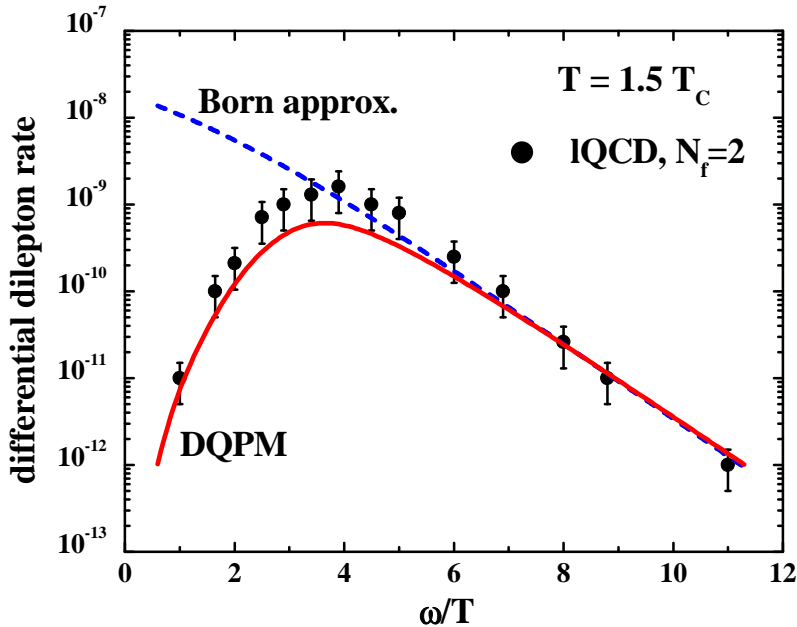}
    \caption{The 'back-to-back' dilepton emission rate (\ref{leptons2}) from the
    DQPM (solid red line) in comparison to the Born approximation
    (\ref{leptons1}) for massless partons (dashed blue line) and the results
    from the lQCD analysis in Ref. \cite{Karsch7} at $T= 1.5\
    T_c$ in case of $N_f=2$ (full dots).}
    \label{fig12}
  \end{center}
\end{figure}

In order to provide some further comparison to lQCD results we
consider the dilepton production rate in thermal equilibrium (at
temperature $T$) in a two-flavor QGP as in Ref. \cite{Karsch7}.
For massless quarks and antiquarks the emission rate of
'back-to-back' leptons ($e^+ e^-$) is given by
\cite{Karsch7,Braten} \be \label{leptons1} \frac{d W}{d \omega d^3
p}({\bf p}=0) = \frac{5 \alpha^2}{36 \pi^4} \
n_F(\frac{\omega}{2T}) \ n_F(\frac{\omega}{2T}) \ , \ee where
$\omega$ is the invariant mass of the lepton pair and $n_F$
denotes the Fermi distribution function. In (\ref{leptons1})
$\alpha$ is the electromagnetic coupling constant.  For massless
partons, furthermore, the magnitude of the lepton momenta is given
by $|{\bf p}| = \omega/2$ while their direction is opposite in the
dilepton rest frame. Neglecting the rest mass of leptons their
energy is $\omega/2$ in the dilepton rest frame, too. The
expression (\ref{leptons1}) changes in case of spectral functions
with finite width to
\be
\label{leptons2} \frac{d W}{d \omega d^3 p} = \frac{5 \alpha^2}{36
\pi^4} \int_{0}^\infty d \omega_1 \ \int_{0}^\infty d \omega_2 \
\int_0^\infty dp \ \frac{\omega_1}{\pi} \frac{\omega_2}{\pi} \
\rho_q(\omega_1,p;T) \rho_{\bar q}(\omega_2,p;T) \ee $$ \cdot
\frac{\sqrt{{\tilde \lambda}(\omega^2,P_1^2,P_2^2)}}{\omega^2} \
\delta(\omega - \omega_1 - \omega_2) \  n_F(\frac{\omega_1}{T}) \
n_F(\frac{\omega_2}{T}) \  $$

\noindent
 with $P_j^2 = \omega_j^2-p^2$ denoting the
invariant mass of the annihilating partons $j=1,2$. In
(\ref{leptons2}) the factor $\sqrt{{\tilde \lambda}}/\omega^2$
gives a flux correction in case of massive quasiparticles with
${\tilde \lambda}(x,y,z) = (x - y -z)^2 - 4 yz$.

Since the spectral functions in the DQPM are fixed (cf. Section 2)
the lepton emission rate (\ref{leptons2}) can be evaluated without
introducing any further assumption (or parameter). The results for
the different emission rate for $N_f = 2$ are shown in Fig.
\ref{fig12} for $T= 1.5 \ T_c$ (solid red line)  in comparison to
the limit (\ref{leptons1}) (dashed blue line) and the lQCD results
from \cite{Karsch7} (full dots). The lQCD dilepton rate has been
obtained from the temporal correlators in lQCD employing the
'maximum entropy method' which has a systematic error in the order
of 30 to 50 \% \cite{Karsch7} depending on the energy scale
considered. The results from Fig. \ref{fig12} demonstrate a
drastic suppression of low mass lepton pairs due to the finite
mass of the partons. On the other hand the spectra from the DQPM
are in qualitative agreement with the lQCD results from Ref.
\cite{Karsch7} when including the systematic uncertainties of the
latter approach. Thus the DQPM (in its present version) passes a
further test in comparison to lQCD.

\section{Conclusions and discussion}
The present study has provided a novel interpretation of the
dynamical quasiparticle model (DQPM) by separating time-like and
space-like quantities for particle densities, energy densities,
entropy densities ect. that also paves the way for an off-shell
transport approach \cite{PHSD}. As known from previous studies
\cite{Andre05,Cassing06} the entropy density $s$ is found to be dominated
 by the on-shell quasiparticle contribution
while the space-like part of the off-shell contribution  gives
only a small (but important) enhancement. However, in case of the
parton 'densities' $N_x = N_x^+ + N_x^-$ and the energy densities
$T_{00,x} = T_{00,x}^+ + T_{00,x}^-$ ($x = g, q, \bar{q}$) the
situation is opposite: here the space-like parts ($N_x^-,
T_{00,x}^-$) dominate over the time-like parts ($N^+, T_{00}^+$)
except close to $T_c$ where the independent quasiparticle limit of
on-shell particles is approximately regained. The latter limit is
a direct consequence of the infrared enhancement of the coupling
(\ref{eq:g2}) close to $T_c$ (in line with the lQCD studies in
Ref. \cite{Bielefeld} ) and a decrease of the width $\gamma$
(\ref{eq:gamma}) when approaching $T_c$ from above.

Since only the time-like part of the parton density can be
propagated within the lightcone the space-like part $N_x^-$ has to
be attributed to $t$-channel exchange partons in scattering
processes that contribute also to the space-like energy densities
$T_{00,x}^-$. The latter quantities may be regarded as potential
energy densities $V_x$. This, in fact, is legitimate since the
total quasiparticle energy density $T^{00}$ (\ref{ent}) very well
matches the energy density (\ref{eps}) obtained from the
thermodynamical relations. Only small deviations indicate that the
DQPM in its straightforward application is not fully consistent in
the thermodynamical sense.

By taking derivatives of the potential energy densities $V_x$ with
respect to the time-like gluon and fermion densities one may
deduce mean-field potentials $U_x$ for gluons and quarks
(antiquarks) as a function of the parton density $\rho_p$ which
enters instead of the thermodynamical Lagrange parameters $T$ and
$\mu_q$. Second derivatives w.r.t. the gluon and/or fermion
densities then define effective interactions between gluons,
quarks and quarks and gluons. We find that the gluon-gluon
interaction is stronger than the quark-quark (antiquark) by
roughly a factor of 4-5 whereas the quark-gluon interaction
strength approximately scales (relative to the quark-quark
interaction) by the ratio of Casimir eigenvalues, i.e. 9/4. The
effective parton-parton interactions are repulsive for $\rho_p
>$ 2.2 fm$^{-3}$ and turn strongly attractive at lower density.
Since at low parton densities (or average distances larger than
0.77 fm) the system may be described by 2- and 3-body dynamics the
strong attraction will lead to color neutral bound states of
gluons (glueballs), quark and antiquarks (mesons) as well as 3
quarks or 3 antiquarks (baryons or antibaryons).
 The mean-field potentials $U_x$ are found to be independent on
the quark chemical potentials $\mu_q$ (within 5\%) such that the
respective analytical approximations (\ref{pott}) are well suited for an
implementation in off-shell parton transport approaches (as e.g.
PHSD).

Since for the case of three light flavors ($N_f$=3) the effective
masses (squared) in the DQPM scale with $T^{*2}:= T^2 + \mu_q^2/\pi^2$ and
the spectral width of the dynamical quasiparticles is assumed to
be given by the 'modified' HTL expression (\ref{gammamu}) an approximate scaling of
various quantities is found in the DQPM at finite quark chemical
potential $\mu_q$. As an example we find that the potential energy
per time-like gluon as well as the potential energy per time-like
fermion are approximately independent on $\mu_q$ which directly
leads to the independence of the mean-fields $U_x$ on $\mu_q$ (as
mentioned above). Furthermore the net quark density is found to
roughly scale as $\rho_q \sim \mu_q T^{*2}$  which might be directly
checked by lQCD. In addition
the energy density $\epsilon$ and the pressure $P$ are found to
increase only slightly with $\mu_q$ close to $T_c$ such that the ratio
$P/\epsilon$ (as a function of $\epsilon$)
is also approximately independent on $\mu_q$.

Explicit comparisons of the DQPM calculations with lQCD results
have been performed for the sound velocity (squared) (\ref{sound})
in the 2+1 flavor case and in the two flavor sector ($N_f$ =2)
 for the net quark density as a function of
temperature $T$ and $\mu_q$ from \cite{Karsch8}. The results are
in an acceptable agreement such that the DQPM - which presently
has been fixed only in the pure Yang-Mills sector ($N_f$ = 0) -
reproduces the dominant dependences on $T$ and $\mu_q$.
Furthermore, it could be shown that the 'back-to-back' dilepton
emission rate from the DQPM in case of $N_f$=2 is in qualitative
agreement with lQCD studies from Ref. \cite{Karsch7} at $T= 1.5 \
T_c$ thus providing a further independent test of the model. Note
that in a conventional quasiparticle model with vanishing width
$\gamma_q$ the dilepton emission rate only gives contributions for
invariant masses above 2 $m_q$ which is not in line with the lQCD
result from Ref. \cite{Karsch7}.

Some note of caution with respect to the present DQPM appears
appropriate: the parameters in the effective coupling
(\ref{eq:g2}) and the width (\ref{eq:gamma}) have been fixed in
the DQPM by the entropy (\ref{sdqp}) to lQCD results for $N_f$=0
assuming the form (\ref{eq:rho}) for the spectral function
$\rho(\omega)$. Alternative assumptions for $\rho(\omega)$ will
lead to slightly different results for the time-like and
space-like densities, energy densities {\it etc.} but not to a
qualitatively different picture. Also it is presently unclear if
the three parameters ($\lambda, T_s/T_c, c$) employed in Section
2.1 are approximately the same in case of two or three dynamical
flavors. More precise calculations from lQCD should allow to put
further constraints on the form of the spectral function
$\rho(\omega)$ and to fix the basic model parameters in the
effective coupling. Also the 'modified' HTL expression
(\ref{gammamu}) for the
quasiparticle width has to be controlled by lQCD at finite $\mu_q$
as well as transport coefficients like the shear viscosity $\eta$
or related correlators.

\vspace{0.5cm} The author acknowledges valuable discussions with
E. L. Bratkovskaya, M. H. Thoma and A. Peshier.


\end{document}